\documentclass[aps,prd,preprint,eqsecnum,nofootinbib,groupedaddress,showpacs,floatfix]{revtex4}
\usepackage[dvips]{graphicx}
\usepackage{mciteplus}
\usepackage{amssymb}
\newif\ifdraft
\drafttrue
\newif\ifpreprint
\preprinttrue

\def\nn{\nonumber}

\def\fig#1{fig.~{\ref{#1}}}

\def\Sect#1{Section~{\ref{#1}}}

\def\eqn#1{eq.~(\ref{#1})}

\def\eqns#1#2{eqs.~(\ref{#1}) and~(\ref{#2})}

\def\e{\epsilon}
\def\eps{\epsilon}

\def\Ord{{\cal O}}
\def\Res{\mathop{\rm Res}}

\def\spa#1.#2{\left\langle#1\,#2\right\rangle}
\def\spb#1.#2{\left[#1\,#2\right]}
\def\spash#1.#2{\spa{\smash{#1}}.{\smash{#2}}}
\def\spbsh#1.#2{\spb{\smash{#1}}.{\smash{#2}}}
\def\sand#1.#2.#3{%
\left\langle\smash{#1^{-}}{\vphantom1}\right|{#2}%
\left|\smash{#3^{-}}{\vphantom1}\right\rangle}
\def\sandp#1.#2.#3{%
\left\langle\smash{#1^{-}}{\vphantom1}\right|{#2}%
\left|\smash{#3^{+}}{\vphantom1}\right\rangle}
\def\sandpp#1.#2.#3{%
\left\langle\smash{#1^{+}}{\vphantom1}\right|{#2}%
\left|\smash{#3^{+}}{\vphantom1}\right\rangle}
\def\sandpm#1.#2.#3{%
\left\langle\smash{#1^{+}}{\vphantom1}\right|{#2}%
\left|\smash{#3^{-}}{\vphantom1}\right\rangle}
\def\sandmp#1.#2.#3{%
   \left\langle\smash{#1^{-}}{\vphantom1}\right|{#2}%
    \left|\smash{#3^{+}}{\vphantom1}\right\rangle}

\newbox\charbox
\newbox\slabox
\def\s#1{{      
        \setbox\charbox=\hbox{$#1$}
        \setbox\slabox=\hbox{$/$}
        \dimen\charbox=\ht\slabox
        \advance\dimen\charbox by -\dp\slabox
        \advance\dimen\charbox by -\ht\charbox
        \advance\dimen\charbox by \dp\charbox
        \divide\dimen\charbox by 2
        \raise-\dimen\charbox\hbox to \wd\charbox{\hss/\hss}
        \llap{$#1$}
}}

\def\tree{{\rm tree}}
\def\oneloop{{1\mbox{-}\rm loop}}
\def\twoloop{{2\mbox{-}\rm loop}}

\begin{document}

\hbox{
Saclay IPhT--T11/208$\hskip 0.55cm \null$
UUITP-21/11 
}

\title{
Maximal Unitarity at Two Loops}

\author{David~A.~Kosower}
\affiliation{Institut de Physique Th\'eorique, CEA--Saclay,
          F--91191 Gif-sur-Yvette cedex, France\\
{\tt David.Kosower@cea.fr}}
\author{Kasper J.~Larsen}
\affiliation{Department of Physics and Astronomy, Uppsala
University, SE--75108 Uppsala, Sweden} \affiliation{Institut de
Physique Th\'eorique, CEA--Saclay,
          F--91191 Gif-sur-Yvette cedex, France\\
{\tt Kasper.Larsen@cea.fr}}

\begin{abstract}
We show how to compute the coefficients of the double box basis
integrals in a massless four-point amplitude in terms
of tree amplitudes.  We show how to choose
suitable multidimensional contours for performing the required cuts, and derive
consistency equations from the requirement that integrals
of total derivatives vanish.
Our formul\ae{} for the coefficients can
be used either analytically or numerically.
\end{abstract}


\maketitle

\section{Introduction}

The computation of higher-order corrections to amplitudes in gauge
theories is important to searches for new physics at modern particle
colliders.  Next-to-leading order (NLO) corrections in quantum
chromodynamics (QCD), in particular, play an important role in
providing a reliable quantitative estimate of backgrounds to possible
signals of new physics~\cite{LesHouches}.  NLO corrections to differential
cross sections require several ingredients
beyond the tree-level amplitudes for the basic process under study:
real-emission corrections, with an additional emitted gluon, or a
gluon splitting into a quark--antiquark pair; and virtual one-loop
corrections, with a virtual gluon or virtual quark in a closed loop.
The required one-loop corrections are challenging with traditional
Feynman-diagram methods, and become considerably more difficult as the
number of final-state partons (gluons or quarks) grows.

The unitarity
method~\cite{UnitarityMethod,Bern:1995db,Zqqgg,DdimensionalI,BCFUnitarity,OtherUnitarity,Bootstrap,BCFCutConstructible,BMST,OPP,OnShellReview,Forde,Badger,DdimensionalII,BFMassive,BergerFordeReview,Bern:2010qa},
a new method which has emerged over the last decade and a half, has
rendered such computations tractable.  It has made possible a variety
of computations of one-loop amplitudes, in particular of processes
with many partons in the final state.  In its most recent form, the
method can be applied either analytically or purely
numerically~\cite{EGK,BlackHatI,CutTools,MOPP,Rocket,BlackHatII,CutToolsHelac,Samurai,WPlus4,NGluon,MadLoop}.
The numerical formalisms underly recent software libraries and
programs that are being applied to LHC phenomenology.  In the current
formalism, the one-loop amplitude in QCD is written as a sum over a
set of basis integrals, with coefficients that are rational in
external spinors,
\begin{equation}
{\rm Amplitude} = \sum_{j\in {\rm Basis}} 
  {\rm coefficient}_j {\rm Integral}_j + 
{\rm Rational}\,.
\label{BasicEquation}
\end{equation}
The integral basis for amplitudes with massless internal lines contains
box, triangle, and bubble integrals in addition to purely rational
terms (dropping all terms of $\Ord(\eps)$ in the dimensional regulator).
  The coefficients
are calculated from products of tree amplitudes, typically by
performing contour integrals.

For NLO corrections to some processes, one-loop amplitudes do not
suffice.  This is the case for subprocesses whose leading-order
amplitude begins at one loop.  An example is the gluon fusion to
diphoton subprocess, $gg\rightarrow\gamma\gamma$, which is an
important background to searches for the Higgs boson at the LHC.
Although this subprocess is nominally suppressed by a power of the
strong coupling $\alpha_s$, the large gluon parton density at smaller
$x$ can compensate for this additional power, giving rise to
contributions to cross sections which are comparable to those from
tree-level quark-initiated
subprocesses~\cite{Ellis:1987xu,Berger:1983yi,Aurenche:1985yk}.  Other
examples include production of electroweak boson pairs, $gg\rightarrow
Z\gamma,ZZ,W^+W^-$.  NLO corrections to such processes at the LHC
require the computation of two-loop amplitudes~\cite{Bern:2002jx}.

Two-loop amplitudes are also required for any studies beyond NLO.
Next-to-next-leading order (NNLO) fixed-order calculations form the
next frontier.  The only existing fully-exclusive NNLO jet
calculations to date are for three-jet production in
electron--positron annihilation~\cite{NNLOThreeJet}.  These are
necessary to determine $\alpha_s$ to 1\% accuracy from jet data at
LEP~\cite{ThreeJetAlphaS}, competitively with other determinations.
At the LHC, NNLO calculations will be useful for determining an honest
theoretical uncertainty estimate on NLO calculations, for assessing scale
stability in multi-scale processes such as $W$+multi-jet production,
and will also be required for precision measurements of new physics once it
is discovered.

The unitarity method has already been applied to higher-loop 
amplitudes.
At one loop, there are different variants of the method.
The basic unitarity approach forms a discontinuity 
out of the product of two tree amplitudes.  
Isolating the coefficients
of specific basis integrals usually still requires performing 
symbolic algebra on
the product of trees; this is not well-suited to a numerical approach, and
also reduces efficiency of an analytic calculation.  Basic unitarity
corresponds to cutting two propagators in a one-loop amplitude.  
Generalized unitarity 
cuts more than two propagators at once, isolating fewer integrals.
`Maximal' generalized unitarity cuts as many propagators as 
possible; in combination with contour integrals over remaining
degrees of freedom, this isolates individual integrals.
At higher loops, `minimal' generalized unitarity cuts the minimum number
of propagators needed to break all loops into a product of trees.
Each cut is again a product of tree amplitudes, but because not
all possible propagators are cut, each generalized cut will
correspond to several integrals and their coefficients, and
algebra will again be required to isolate specific integrals and
their coefficients.  This approach does have the advantage
of not requiring a basis of integrals.
A number of calculations have been done this way, primarily
in the ${\cal N}=4$ supersymmetric gauge theory
and ${\cal N}=8$ supergravity~\cite{Bern:1997nh,ABDK,
Bern:2005iz,Bern:2006vw,Bern:2006ew,Bern:2007ct,Bern:2008ap,Bern:2008pv,LeadingSingularity,ArkaniHamed:2009dn,ArkaniHamed:2010kv,Bern:2010tq,Kosower:2010yk,ArkaniHamed:2010gh,Carrasco:2011mn,Carrasco:2011hw,Bern:2011rj,LarsenSixPoint}, 
but including
several four-point calculations in QCD
and supersymmetric
theories with less-than-maximal supersymmetry~\cite{Bern:2000dn,Bern:2002tk,BernDeFreitasDixonTwoPhoton,Bern:2002zk,Bern:2003ck,TwoLoopSplitting,DeFreitas:2004tk}.

In this paper, we take the first steps in developing the 
maximal generalized unitarity approach at two loops in a form
suitable for both analytic and numerical calculation.  We show
how to extract the coefficient of the planar double box to
leading order in the dimensional regulator $\eps$.  Higher-loop
amplitudes can be written in a similar form to those at 
one loop~(\ref{BasicEquation}),
as a sum over an integral basis~\cite{TwoLoopBasis}, 
along with possible rational terms.
At higher loops, however, the coefficients of the basis
integrals are no longer functions
of the external spinors alone, but will depend explicitly on $\eps$.
Just as at one loop, computing coefficients requires choosing contours
for the unfrozen degrees of freedom.  We use the equations relating
generic tensor integrals to basis or master integrals in order to ensure the
consistency and completeness of the choice of contours.  The extraction
of the double-box coefficient bears a superficial similarity 
to the procedure that would
be followed in the leading-singularity 
approach~\cite{Octacut,LeadingSingularity}, 
but unlike the latter, manifestly ensures the consistency of
the extraction with respect to terms that vanish after integration.
Such terms inevitably arise when using the integration-by-parts (IBP)
approach~\cite{IBP,Laporta,GehrmannRemiddi,LIdependent,AIR,FIRE,Reduze,SmirnovPetukhov}
 in relating formally-irreducible tensor integrals to basis
integrals.
The extraction of higher-order terms in $\eps$ or the coefficients
of integrals with fewer propagators, both of which we leave to
future work, would also be different.

During the preparation of this manuscript, a preprint by
Mastrolia and Ossola appeared~\cite{MastroliaOssola}, 
analyzing the two-loop integrand in a generalization of
the formalism of Ossola, Papadopoulos and Pittau (OPP)~\cite{OPP},
following a complementary approach
to unitarity at two loops.

This paper is organized as follows.  In \Sect{ReviewSection}, we
review maximal generalized unitarity at one loop, focusing on the
computation of the coefficients of the box integral.  In
\Sect{ParametrizationSection}, we give an outline of the two-loop
formalism, and detail the solutions to the cut equations.  In
\Sect{ConstraintSection}, we present the set of constraint equations,
and their solutions.  In \Sect{ExtractionSection}, we give the master
formul\ae{} for the double-box coefficients, and give some examples of
their use in \Sect{ExamplesSection}.  We summarize in
\Sect{ConclusionSection}.

\section{Maximal Unitarity at One Loop}
\label{ReviewSection}

We begin by reviewing the derivation of the formula for coefficients of
one-loop boxes using quadruple cuts, originally written down by Britto,
Cachazo, and Feng~\cite{BCFUnitarity}.  
We adopt an approach and notation that generalize to our derivation
for two-loop coefficients in following sections.
Our starting point is the formal
diagrammatic expression for the amplitude,
\begin{equation}
{\rm Amplitude} = \sum_{{\rm Feynman\atop\rm diagrams} F} 
\int {d^D\ell\over(2\pi)^D}\;\mathop{\rm Numerator}\nolimits_{F}(\ell,\cdots)
\,\cdot\,\mathop{\rm Propagators}\nolimits_{F}(\ell,\cdots)\,,
\label{StartingEquation}
\end{equation}
where the ellipses represent dependence on external momenta, polarization
vectors, and spinor strings.  Although the whole point of the method is to
avoid computing any Feynman diagrams explicitly, it is still convenient to
refer to them in the abstract, as a means of providing the connection to
field theory and to Feynman integrals.

Applying tensor and integral
reductions~\cite{IntegralReductions}, along with a
Gram-determinant identity holding through $\Ord(\eps^0)$, we obtain the
basic equation~(\ref{BasicEquation}) without any reference to unitarity or
on-shell conditions.  (In a slight abuse of language, we will refer
to integrals with no free indices, but numerator powers of the loop momentum
contracted into external vectors, as ``tensor integrals''.)

At one loop, it is sufficient for our purposes to concentrate 
on the four-dimensional components of the loop momentum.
  (The accompanying integrals must of course be
evaluated keeping the full $(D=4-2\eps)$-dimensional dependence.)
In order to derive formul\ae{} for the coefficients of basis
integrals, we apply
cuts to both sides of \eqn{StartingEquation}.  In the basic
unitarity method, we would replace two propagators, separated by
a non-null sum of external momenta, by delta functions which
freeze the loop momenta they carry to their on-shell values.  In generalized 
unitarity~\cite{Zqqgg,BCFUnitarity}, 
we would like to apply additional delta functions to
put additional momenta to on-shell values.  However, once we put
the momenta carried by more than
two massless propagators to their on-shell valuess, the solutions to
the on-shell equations are complex, and taken at face value, 
the delta functions
would actually yield zero.

The same issue arose in the evaluation of the connected 
prescription~\cite{RSV}
for amplitudes in Witten's twistor string theory~\cite{WittenTwistorString};
 the solution is to use contour integrals instead of delta 
functions~\cite{Vergu,ArkaniHamed:2009dn,Bullimore:2009cb}.  
To do so, we think
of complexifying the space in which the four-dimensional loop
momenta live, from $\mathbb{R}^{1,3}$ to $\mathbb{C}^4$, and taking the
integrals on both sides of \eqn{BasicEquation} to be over a product of
contours running along the real axis.  
We can imagine evaluating the loop integrals along other contours
as well.
New contours that will be useful for our purposes are those whose
product encircles simultaneous poles in all four-dimensional components of
the loop momentum.  Performing the four-dimensional loop-momentum integral
over each such contour will yield the residue at the corresponding
encircled joint or global pole.  The residue extracts~\cite{} the terms in the
integrand which contain each of the corresponding propagators,
removes the denominators, divides by the appropriate Jacobian, and sets the
components of the loop momentum to their values at the joint pole.

The Jacobian is a determinant which
arises from the transformation to variables which
express each denominator factor linearly in a different variable.
Unlike a product of delta functions, which would produce a
factor of the inverse of the absolute value of the Jacobian,
the transformation here will produce a factor of the inverse
of the Jacobian.  This ensures that the factor is analytic in
any variables on which it depends, so that further contour
integrations can be carried out.

\def\QuadCutTorus{T_Q}
Notationally, it will still be convenient to use delta functions;
to do so, define the product of delta functions to yield exactly this
contour integral,
\begin{eqnarray}
&&\int {d^4\ell\over (2\pi)^4}\; {\rm Numerator}_F(\ell,\cdots)
\delta\bigl(\ell^2\bigr)\delta\bigl( (\ell-k_1)^2\bigr)
\delta\bigl( (\ell-k_1-k_2)^2\bigr) \delta\bigl( (\ell+k_4)^2\bigr)
\equiv \nn\\
&&\hskip 10mm
\oint_{\QuadCutTorus} 
{d^4\ell\over (2\pi)^4}\; {{\rm Numerator}_F(\ell,\cdots)
   \over \ell^2 (\ell-k_1)^2 (\ell-k_1-k_2)^2(\ell+k_4)^2}\,,
\end{eqnarray}
where we have divided out overall factors of $2\pi i$ associated
with each delta function, and where $\QuadCutTorus$ is a four-torus
encircling the solutions to the simultaneous equations,
\begin{equation}
\ell^2 = 0\,,\quad
(\ell-k_1)^2 = 0\,,\quad
(\ell-k_1-k_2)^2 = 0\,,\quad
(\ell+k_4)^2 = 0\,,\quad
\label{OneLoopQuadCut}
\end{equation}
and where --- in a non-standard bit of notation --- we absorb
a factor of $1/(2\pi i)$ into the definition of each contour integral,
so that evaluating the four-fold contour integral
yields a sum over residues with no additional
factors of $2\pi i$.

In four-point amplitudes, the external momenta do not suffice
to provide a basis for arbitrary external vectors; to three of
them (say $k_1$, $k_2$, and $k_4$), we need to add another
external vector, for example $v^\mu=\varepsilon(\mu,k_1,k_2,k_4)$.
Then we can express all dot products
of the loop momentum with external vectors in terms of four dot products:
$\ell\cdot k_1$, $\ell\cdot k_2$, $\ell\cdot k_4$, and
$\ell\cdot v$.  In reducing integrals, 
odd powers of $v\cdot\ell$ will give rise to vanishing integrals because
of parity; even
powers can be re-expressed in terms of Gram determinants 
and thence in terms of the other dot products
(up to terms involving the $(-2\eps)$-dimensional components of
the loop momentum).  The remaining three dot products can be
re-expressed as linear combinations of propagator denominators
and external invariants, allowing integrals with insertions of
them in the numerator to be simplified.

One would be tempted to believe that replacing the original contours
running along the real axis with some other contour, such as
$\QuadCutTorus$, would leave the equality~(\ref{BasicEquation})
undisturbed, but this is not quite right, because there are implicitly
terms in the integrand of the left-hand side that are `total
derivatives', that is terms which integrate to zero.  These terms
arise during the integral reductions described above.  They were made
explicit in the decomposition of the integrand used by OPP~\cite{OPP}.
For general contours, the reduction equation will then take the form,
\begin{equation}
{\rm Amplitude} = \sum_{j\in {\rm Basis}} c_j I_j + 
{\rm Rat} 
+ \sum_{j\in {\rm Basis}} \sum_{t\in {\rm Total\atop\rm Derivatives}} 
  c'_{j,t} U_{j,t}\,,
\label{GeneralContourReduction}
\end{equation}
where each $U_{j,t}$ is the integral of an expression $W_{j,t}$ which
would vanish if taken over the real slice, for example $W_{1,1} =
\varepsilon(\ell,k_1,k_2,k_4)$:
\begin{equation}
U_1 = \int {d^D\ell\over (2\pi)^D}\; 
  {\varepsilon(\ell,k_1,k_2,k_4)
   \over \ell^2 (\ell-k_1)^2 (\ell-k_1-k_2)^2(\ell+k_4)^2}\,.
\label{TotalDerivativeExample}
\end{equation}
This integral will no longer necessarily vanish if we integrate over general 
contours in $\mathbb{C}^4$.  In this equation,
\begin{equation}
\varepsilon(q_1,q_2,q_3,q_4) \equiv \varepsilon_{\mu\nu\lambda\sigma}
q_1^\mu q_2^\nu q_3^\lambda q_4^\sigma\,,
\end{equation}
where $\varepsilon_{\mu\nu\lambda\sigma}$ is the standard
antisymmetric Levi-Civita tensor. Another example is the cube
of the Levi-Civita symbol, $W_{1,2} = W_{1,1}^3$, though as it turns
out, this latter integrand does not contribute any additional equations
below.

\begin{figure}[!h]
\begin{center}
\includegraphics[angle=0, width=0.32\textwidth]{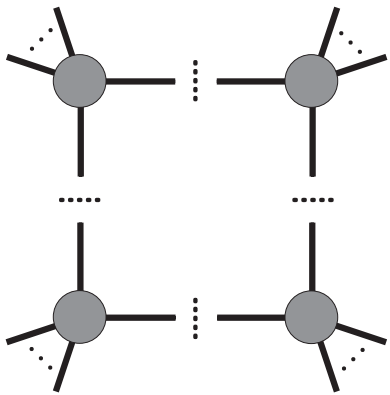}
\end{center}
\caption{The general quadruple cut in a one-loop box.}
\label{OneLoopGeneralBoxFigure}
\end{figure}

\def\boxn{\mathrm{box}}
When we perform a quadruple cut, that is the integral over $\QuadCutTorus$,
we will restrict the set of Feynman diagrams in the expression for
the amplitude to those containing all four
propagators; cut the propagators; and impose the on-shell conditions
corresponding to the vanishing of the propagator denominators.  If we
imagine working in a physical gauge (such as light-cone gauge), this also
restricts the cut lines to have physical polarizations.  Each diagram then
falls apart into a product of four diagrams, one corresponding to each
corner of the box.  The sum over diagrams factorizes into a four-fold sum
over the tree diagrams at each corner, as shown in
\fig{OneLoopGeneralBoxFigure}.  Each such sum will give an on-shell tree
amplitude, with two cut loop momenta and the external legs attached to that
corner as arguments,
\begin{eqnarray}
{\rm A}^{\oneloop} &=&
\sum_{{\rm Feynman\atop\rm diagrams} F} \int {d^D\ell\over(2\pi)^D}\;
{N_F\over D_F}\nn\\
&\rightarrow&
\sum_{{\rm helicities\atop\rm species}} 
  \sum_{{\rm Feynman\atop\rm diagrams} A}  {N_A\over D_A}
   \sum_{{\rm Feynman\atop\rm diagrams} B}  {N_B\over D_B}
   \sum_{{\rm Feynman\atop\rm diagrams} C}  {N_C\over D_C}
   \sum_{{\rm Feynman\atop\rm diagrams} D}  {N_D\over D_D}\\
&=& \sum_{{\rm helicities\atop\rm species}} 
A^\tree_A A^\tree_B A^\tree_C A^\tree_D\,.\nn
\end{eqnarray}
  (If we had not initially used a physical gauge, it is
at this stage, summing over all diagrams,
 that we would recover the restriction to physical
polarizations.)  Integrating over $\QuadCutTorus$
in \eqn{GeneralContourReduction}
will give us the following equation,
\begin{equation}
\sum_{{\rm helicities\atop\rm species}} 
A^\tree_A A^\tree_B A^\tree_C A^\tree_D = 
c_\boxn + \sum_t c_{\boxn,t}' U_{\boxn,t}
\end{equation}
where the Jacobian has canceled out of the equality,
and the sum on the right-hand side
runs over possible total derivatives with a box
integrand.  In order
to solve for the desired coefficient $c_\boxn$, or equivalently
to ensure that the equality in
\eqn{BasicEquation} is maintained, we must evaluate
the integral over
a linear combination of new contours such that all possible
integrals of 
`total derivatives' $U_t$ are projected out.  
As we will show later, this requirement
gives us constraints
that determine the allowed combinations of contours, and in turn
the equations for the coefficients of the box integrals.

In the case of the one-loop box integral, the joint-pole equations
are given by \eqn{OneLoopQuadCut}
or equivalently by,
\begin{equation}
\ell^2 = 0\,,\quad
\ell\cdot k_1 = 0\,,\quad
\ell\cdot k_2 = {s\over2}\,,\quad
\ell\cdot k_4 = 0\,,\quad
\label{OneLoopQuadCut2}
\end{equation}
which form makes it clear that there are two distinct solutions,
and hence two distinct contours.  The one `total derivative' we must consider
is the $\varepsilon$ expression $U_1$ given above 
in \eqn{TotalDerivativeExample}.  
It turns out that it
evaluates to compensating values on the two solutions, so that summing
over them projects it out, and hence gives an equation for the 
coefficient of the box in terms of the product of tree amplitudes at
each corner,
\begin{equation}
c_\boxn = {1\over2} \sum_{{\rm helicities\atop\rm species}} 
       A^\tree_A A^\tree_B A^\tree_C A^\tree_D\,.
\label{QuadCutCoeff}
\end{equation}

As an example, study the coefficient of the one-mass box with
$m_3^2\neq 0$.  Parametrize the four-dimensional part of the loop
momentum as follows,
\begin{equation}
\ell^\mu = \alpha_1 k_1^\mu + \alpha_2 k_2^\mu 
           +{\alpha_3 s\over2\spa1.4\spb4.2} \sand1.{\gamma^\mu}.2
           +{\alpha_4 s\over2\spa2.4\spb4.1} \sand2.{\gamma^\mu}.1\,.
\label{OneLoopParametrization}
\end{equation}
A general contour integral for the four-dimensional
part of the box integral then takes the form,
\begin{eqnarray}
{1\over s^4} \int_{C} {d^4\alpha_i}\; J_\alpha&&
  {1\over (\alpha_1\alpha_2-\omega \alpha_3\alpha_4)
   (\alpha_1\alpha_2-\omega \alpha_3\alpha_4-\alpha_2)
(\alpha_1\alpha_2-\omega \alpha_3\alpha_4-\alpha_2-\alpha_1+1)}\nn\\
&& \times {1\over (\alpha_1\alpha_2-\omega \alpha_3\alpha_4
+\alpha_1 t/s + \alpha_2 u/s + \alpha_3 +\alpha_4)}
\label{GeneralOneLoopContourIntegral}
\end{eqnarray}
where $\omega = s^2/(t u)$.  
In this expression, $J_\alpha$
is the Jacobian that arises from changing variables from the $\ell^\mu$ to
the $\alpha_i$. (We do not need its explicit form, only the knowledge
that it is independent of the $\alpha_i$, a consequence of the linearity
of the $\ell^\mu$ in the $\alpha_i$.)

\def\Sol{{\cal S}}
The cut equations~(\ref{OneLoopQuadCut2}) then take the form,
\begin{eqnarray}
\biggl(\alpha_1 \alpha_2 - {s^2\over t u} \alpha_3\alpha_4\biggr) s &=& 0\,,\nn\\
\alpha_2 s &=& 0\,,\nn\\
\alpha_1 s &=& s\,,\\
\alpha_1 t + \alpha_2 u + \alpha_3 s+\alpha_4 s &=& 0\,,\nn
\end{eqnarray}
which have two solutions,
\begin{eqnarray}
\Sol_1: &&\alpha_1 = 1\,,\quad
\alpha_2 = 0\,,\quad
\alpha_3 = -{t\over s}\,,\quad
\alpha_4 = 0\,;\nn\\
\Sol_2: &&\alpha_1 = 1\,,\quad
\alpha_2 = 0\,,\quad
\alpha_3 = 0\,,\quad
\alpha_4 = -{t\over s}\,.
\end{eqnarray}
If we define $C_j(v)$ to be a small circle in the 
complex $\alpha_j$-plane
that encloses the point $v$, then the two contours we must consider
are,
\begin{eqnarray}
T_1 &=&C_1(1) \times C_2(0) \times C_3(-t/s)\times C_4(0)\,;\nn\\
{\rm and\ \ } T_2 &=& C_1(1) \times C_2(0) \times C_3(0)\times C_4(-t/s)\,.
\end{eqnarray}

We can evaluate the four-fold
integral~(\ref{GeneralOneLoopContourIntegral}) by `global
residues'~\cite{Shabat,ArkaniHamed:2009dn}.  The sign
of the result will depend on the orientation chosen for the
contour; but this sign will drop out of final formul\ae{} for
integral coefficients so long as this orientation is chosen
consistently throughout the calculation.
To do so, we
should first change to variables where each pole is in a different
variable, and where the denominators are linear in that variable with
unit coefficient.  The Jacobian from this change of variables will
take the form,
\begin{equation}
J_1 = \det_{i,j}\biggl({\partial f_j\over\partial\alpha_i}\biggr)\,,
\end{equation}
where
\begin{eqnarray}
f_1 &=&  \alpha_1\alpha_2-\omega \alpha_3\alpha_4\,,\nn\\
f_2 &=&   \alpha_1\alpha_2-\omega \alpha_3\alpha_4-\alpha_2\,,\nn\\
f_3 &=& \alpha_1\alpha_2-\omega \alpha_3\alpha_4-\alpha_2-\alpha_1+1\,,\\
f_4 &=& \alpha_1\alpha_2-\omega \alpha_3\alpha_4
+\alpha_1 t/s + \alpha_2 u/s + \alpha_3 +\alpha_4\,.
\end{eqnarray}
We find,
\begin{equation}
J_1 = \omega (\alpha_4-\alpha_3)\,.
\end{equation}
Evaluating the box integral with a numerator ${\rm Num}(\ell,\cdots)$
along a contour given by a linear combination of the two $T_i$ with
weights $a_i$, we obtain,
\begin{equation}
s^{-4} J_\alpha \left(a_1 J_1^{-1} {\rm Num}(\ell,\cdots) \Big|_{\Sol_1}
+a_2 J_1^{-1} {\rm Num}(\ell,\cdots) \Big|_{\Sol_2}\right)\,.
\end{equation}
Using the parametrization~(\ref{OneLoopParametrization}), we find
the following expression for the Levi-Civita symbol we need,
\begin{eqnarray}
\varepsilon(\ell,k_1,k_2,k_4) &=&
\alpha_3 s \varepsilon\left(
 {\textstyle \frac{\sand1.{\gamma^\mu}.2}{2\spa1.4\spb4.2}},k_1,k_2,k_4\right)
+\alpha_4 s\varepsilon\left(
 {\textstyle \frac{\sand2.{\gamma^\mu}.1}{2\spa2.4\spb4.1}},k_1,k_2,k_4\right)
\nn\\
&=& s (\alpha_3-\alpha_4)\varepsilon\left(
 {\textstyle \frac{\sand1.{\gamma^\mu}.2}{2\spa1.4\spb4.2}},k_1,k_2,k_4\right)
\,.
\end{eqnarray}
The constraint that $U_1 = 0$ on the quadruple cut then implies that,
\begin{equation}
-s^{-4} J_\alpha \omega^{-1} \bigl( a_1 + a_2 \bigr) = 0\,.
\label{OneLoopConstraint}
\end{equation}
so that $a_2 = -a_1$.  Higher odd powers of the Levi-Civita tensor lead to the
same constraint.

If we evaluate both sides of \eqn{BasicEquation} on the linear
combination of contours, we find,
\begin{equation}
s^{-4} J_\alpha \sum_{i=1}^2 a_i J_1^{-1} \prod_{j=1}^4 A^\tree_j\Big|_{\Sol_i} 
= s^{-4} J_\alpha  c
 \left( a_1 J_1^{-1}\Big|_{\Sol_1} + a_2 J_1^{-1}\Big|_{\Sol_2}\right)\,,
\end{equation}
where the product is over the tree amplitudes associated to each of the
four vertices of the quadruple-cut box integral in
\fig{OneLoopGeneralBoxFigure}.  Substituting in
the solution to \eqn{OneLoopConstraint}, we find for the
coefficient of the one-loop box,
\begin{equation}
c_\boxn = \frac12 \sum_{i=1}^2 \prod_{j=1}^4 A^\tree_j\Big|_{\Sol_i}
\end{equation}
which is just \eqn{QuadCutCoeff} when summed over possible helicity
assignments and species circulating in the loop.

In the following sections, we show how to generalize these considerations
to two loops.

\section{Maximal Cuts at Two Loops}
\label{ParametrizationSection}

Our basic approach to the planar double box at two loops will be
similar to that reviewed above at one loop.  We use a convenient
parametrization of the loop momenta and choose new contours of
integration to freeze the momenta flowing through all propagators.  We
choose those contours so that constraint equations arising from
consistency conditions are satisfied.  Unlike the procedure at one
loop, cutting all seven propagators does not freeze all components of
both loop momenta, so we must choose new contours for the remaining
unfrozen degrees of freedom as well.  In addition, we have a much
larger and richer set of consistency conditions arising from IBP
identities.  Once we have solved the constraint equations, we will
solve for the coefficients of specific basis integrals.

\def\DBox{P^{**}_{2,2}}

\begin{figure}[!h]
\begin{center}
\includegraphics[angle=0, width=0.4\textwidth]{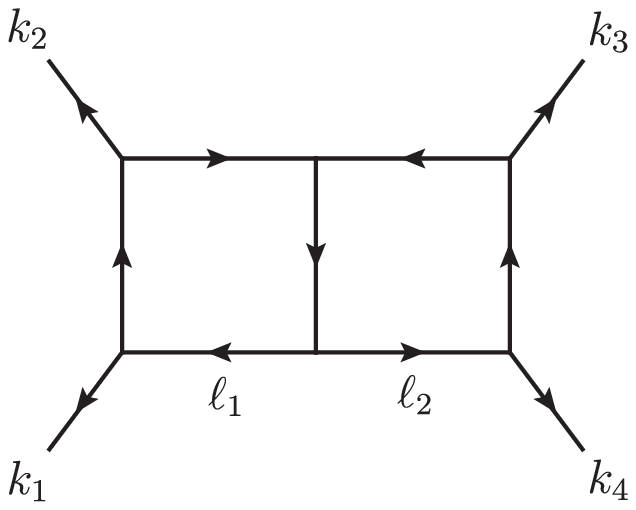}
\end{center}
\caption{The double box integral $\DBox$.}
\label{PlanarTwoLoopDoubleBoxFigure}
\end{figure}

In this section, we give a convenient parametrization of the loop
momenta, and use it to solve on-shell equations.  We list these
solutions below, along with the poles and possible contours for the
remaining unfrozen degrees of freedom.

The two-loop double box integral, shown in~\fig{PlanarTwoLoopDoubleBoxFigure},
 is,
\begin{equation}
\DBox = 
\int {d^{D} \ell_1\over (2\pi)^D} \hspace{0.5mm} {d^{D} \ell_2\over(2\pi)^D}
\hspace{0.7mm} \frac{1}{\ell_1^2(\ell_1 - k_1)^2
(\ell_1 - K_{12})^2(\ell_1 + \ell_2)^2\ell_2^2
(\ell_2 - k_4)^2(\ell_2 - K_{34})^2} \, ,
\label{eq:double_box_integral_def}
\end{equation}
where $K_{i\cdots j} \equiv k_i+\cdots+k_j$, and the notation
follows ref.~\cite{TwoLoopBasis}.

We will focus in this paper on extracting coefficients 
of basis integrals
only to
leading order in the dimensional regulator $\e$, for which it suffices
to consider the four-dimensional components of the loop momentum as
far as cuts are concerned.  The double box has seven propagators; if
we cut all of them, that is put all of the momenta they are
carrying to on-shell values, we will be left with
one additional degree of freedom.  To cut the momenta in this way,
we must shift the contours of integration for the components of 
the two loop momenta $\ell_1$
and $\ell_2$ to encircle the joint solutions to the on-shell equations,
\begin{eqnarray}
&&\ell_1^2 = 0\,,\quad
(\ell_1-k_1)^2 = 0\,,\quad
(\ell_1-K_{12})^2 = 0\,,\quad
\ell_2^2 = 0\,,\quad
(\ell_2-k_4)^2 = 0\,,\quad\nn\\
&&(\ell_2-K_{34})^2 = 0\,,
(\ell_1+\ell_2)^2 = 0\,.
\label{HeptacutEquations}
\end{eqnarray}

\begin{figure}[!h]
\begin{center}
\includegraphics[angle=0, width=0.4\textwidth]{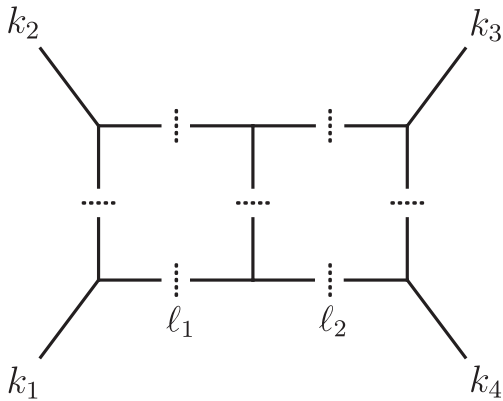}
\end{center}
\caption{The heptacut double box.}
\label{HeptacutDoubleBoxFigure}
\end{figure}

As explained in \Sect{ReviewSection}, we can write the four-dimensional
heptacut integral symbolically as,
\begin{eqnarray}
&&\int {d^{4} \ell_1\over (2\pi)^4} \hspace{0.5mm} {d^{4} \ell_2\over(2\pi)^4}
\hspace{0.7mm}
\delta\big(\ell_1^2\big) \delta\big((\ell_1 - k_1)^2\big)
\delta\big((\ell_1 - K_{12})^2\big) \delta\big((\ell_1 + \ell_2)^2\big)
\nn\\
&&\hskip 10mm\times
\delta\big(\ell_2^2\big) \delta\big((\ell_2 - k_4)^2\big)
\delta\big((\ell_2 - K_{34})^2\big)\,,
\label{HeptacutDoubleBox}
\end{eqnarray}
again dropping overall factors of $2\pi i$ associated with the
delta functions.
This heptacut is depicted in \fig{HeptacutDoubleBoxFigure}.

To solve the on-shell equations, we use the following 
parametrization of the loop momenta,
\begin{eqnarray}
\ell_1^\mu &=& \alpha_1 k_1^\mu + \alpha_2 k_2^\mu + \frac{s_{12}
\alpha_3}{2\spa1.4\spb4.2} \sand1.{\gamma^\mu}.2
+ \frac{s_{12} \alpha_4}{2\spa2.4 \spb4.1}\sand2.{\gamma^\mu}.1
\,,\nn\\
\ell_2^\mu &=& \beta_1 k_3^\mu + \beta_2 k_4^\mu + \frac{s_{12}
\beta_3}{2\spa3.1\spb1.4} \sand3.{\gamma^\mu}.4
 + \frac{s_{12} \beta_4}{2\spa4.1\spb1.3}
\sand4.{\gamma^\mu}.3\,.
\label{TwoLoopParametrization}
\end{eqnarray}

Using this parametrization, 
the six corresponding heptacut equations involving only one loop momentum
are,
\begin{eqnarray}
\ell_1^2 &=& s_{12} \left( \alpha_1 \alpha_2 +
\textstyle{\frac{\alpha_3 \alpha_4}{\chi
(\chi + 1)}}\right)=0\,, \nn \\
(\ell_1 - k_1)^2 &=& s_{12} \left( (\alpha_1 -1) \alpha_2 +
\textstyle{\frac{\alpha_3 \alpha_4}{\chi
(\chi + 1)}}\right) = 0\,, \nn \\
(\ell_1 - K_{12})^2 &=& s_{12} \left( (\alpha_1 -1) (\alpha_2 -1) +
\textstyle{\frac{\alpha_3 \alpha_4}{\chi (\chi + 1)}}\right) = 0\,,
\label{HeptacutEquationsI}\\
\ell_2^2 &=& s_{12} \left( \beta_1 \beta_2 + \textstyle{\frac{\beta_3
\beta_4}{\chi
(\chi +1)}} \right)=0\,, \label{eq:q_squared_parametrized} \nn\\
(\ell_2 - k_4)^2 &=& s_{12} \left( \beta_1 (\beta_2 -1) +
\textstyle{\frac{\beta_3 \beta_4}{\chi
(\chi + 1)}}\right)=0\,, \label{eq:q1_squared_parametrized} \nn\\
(\ell_2 - K_{34})^2 &=& s_{12} \left( (\beta_1 -1) (\beta_2 -1) +
\textstyle{\frac{\beta_3 \beta_4}{\chi (\chi + 1)}} \right)=0\,,\nn
\label{eq:q2_squared_parametrized}
\end{eqnarray}
where
\begin{equation}
\chi \equiv \frac{s_{14}}{s_{12}} \, .
\end{equation}

We can simplify these equations, obtaining
\begin{eqnarray}
\alpha_1 \hspace{-1.5mm}&=&\hspace{-1.5mm} 1 \, , \hspace{5mm}
\alpha_2 \hspace{0.5mm}=\hspace{0.5mm} 0 \, , \hspace{5mm}
\alpha_3 \alpha_4 \hspace{0.5mm}=\hspace{0.5mm} 0\,,
\nn\label{eq:on-shell_values_of_alpha}\\
\beta_1 \hspace{-1.5mm}&=&\hspace{-1.5mm} 0 \, , \hspace{5mm}
\beta_2 \hspace{0.5mm}=\hspace{0.5mm} 1 \, , \hspace{5mm} \beta_3
\beta_4 \hspace{0.5mm}=\hspace{0.5mm} 0\,.
\label{eq:on-shell_values_of_beta}
\label{AlphaBetaEquations}
\end{eqnarray}
These equations have four distinct solutions.  If we substitute these
values into \eqn{HeptacutEquations}, we find
for the last equation,
\begin{eqnarray}
0 &=& (\ell_1 + \ell_2)^2 =
 2 \ell_1 \cdot \ell_2 =\nn\\
&&\hskip 5mm 2 \left( 
k_1^\mu + \frac{s_{12}
\alpha_3}{2\spa1.4\spb4.2} \sand1.{\gamma^\mu}.2
+ \frac{s_{12} \alpha_4}{2\spa2.4 \spb4.1}\sand2.{\gamma^\mu}.1\right)
 \nn \\
&\phantom{=}& \hspace{18mm}\times \left( 
k_{4\mu} + \frac{s_{12}
\beta_3}{2\spa3.1\spb1.4} \sand3.{\gamma_\mu}.4
 + \frac{s_{12} \beta_4}{2\spa4.1\spb1.3}\sand4.{\gamma_\mu}.3\right) \,.
\phantom{aaaaaa}
\label{LastOnShellEquation}
\end{eqnarray}
For two of the four solutions to eqs.~(\ref{AlphaBetaEquations}), this
equation has two solutions, so that overall we find six solutions to
the heptacut equations~(\ref{HeptacutEquationsI},
\ref{LastOnShellEquation}).  To each of the six solutions $\Sol_j$, we
can associate a seven-torus in the parameters $\alpha_i$ and $\beta_i$
that encircles the solution.

For the solution $\alpha_4=0=\beta_4$, the last 
equation~(\ref{LastOnShellEquation}) simplifies to,
\begin{equation}
0 = \left(\spb4.1 + \frac{s_{12}\alpha_3}{\spa1.4}\right)
\left(\spa1.4-\frac{s_{12}\beta_3}{\spb1.4}\right)\,,
\end{equation}
which has two distinct solutions,
\begin{eqnarray}
\Sol_1:&& \alpha_3 = -\chi\,,\quad \beta_3 {\rm\ arbitrary}\,;\nn\\
\Sol_2:&& \beta_3 = -\chi\,,\quad \alpha_3 {\rm\ arbitrary}\,.
\label{Solutions1and2}
\end{eqnarray}
In all solutions, we will change variables so that the remaining
degree of freedom is called $z$.

Likewise, the solution $\alpha_3=0=\beta_3$ also yields two
solutions to \eqn{LastOnShellEquation},
\begin{eqnarray}
\Sol_3:&& \alpha_4 = -\chi\,,\quad \beta_4=z\,;\nn\\
\Sol_4:&& \beta_4 = -\chi\,,\quad \alpha_4 =z\,.
\end{eqnarray}
For the remaining two solutions, the last equation~(\ref{LastOnShellEquation})
 does not factorize,
and we obtain only one solution; for $\alpha_3=0=\beta_4$,
\begin{eqnarray}
\Sol_5:&& \alpha_4 = z\,,\quad \beta_3 = -(\chi+1)\frac{z+\chi}{z+\chi+1}\,;
\label{Solution5}
\end{eqnarray}
and for $\alpha_4 = 0 = \beta_3$,
\begin{eqnarray}
\Sol_6:&& \alpha_3 = z\,,\quad \beta_4 = -(\chi+1)\frac{z+\chi}{z+\chi+1}\,.
\label{Solution6}
\end{eqnarray}
In the last two solutions, we could equally well have chosen a different
parametrization, where $\beta_3$ or $\beta_4$ respectively are set
to $z$.  This just amounts to a change of variables, of course, but
does break the manifest symmetry between the two loops.

\begin{figure}[!ht]
\begin{minipage}[b]{0.45\linewidth}
\begin{center} \includegraphics[scale=0.9]{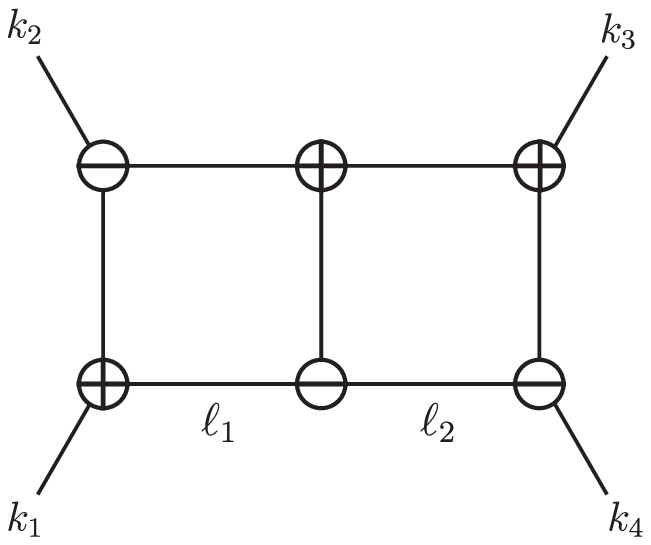} \end{center}
{\renewcommand{\arraystretch}{1.3}
\vspace{-5.1mm}
Solution $\Sol_1$, obtained by setting
\vspace{-5mm}
\begin{eqnarray*}
\begin{array}{ll} 
\smash{\alpha_3 = -\chi}\,, & \hspace{7mm} \smash{\beta_3 = z}\,, \\[-5mm]
\alpha_4 = 0\,, & \hspace{7mm} \beta_4 = 0\,.
\end{array}
\end{eqnarray*}}
\end{minipage}
\begin{minipage}[b]{0.45\linewidth}
\begin{center} \includegraphics[scale=0.9]{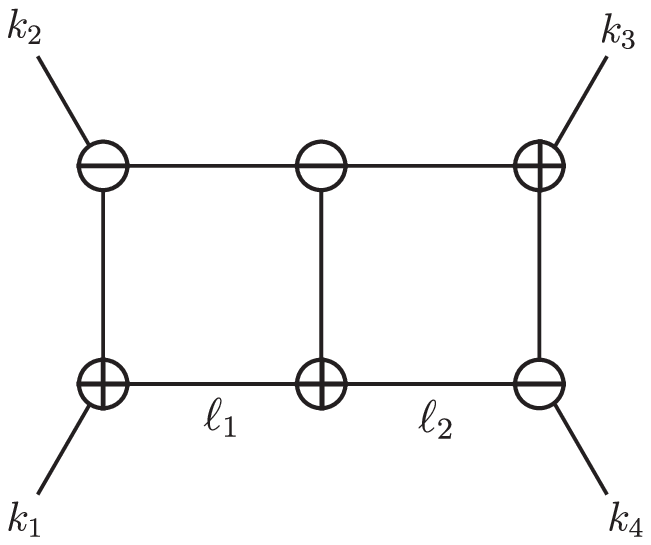} \end{center}
{\renewcommand{\arraystretch}{1.3}
\vspace{-5.1mm}
Solution $\Sol_2$, obtained by setting
\vspace{-5mm}
\begin{eqnarray*}\begin{array}{ll} 
\alpha_3 = z\,, & \hspace{7mm} \beta_3 = -\chi\,, \\[-5mm]
\alpha_4 = 0\,, & \hspace{7mm} \beta_4 = 0\,.
\end{array}
\end{eqnarray*}}
\end{minipage}
%
%
\begin{minipage}[b]{0.45\linewidth}
\begin{center} \includegraphics[scale=0.9]{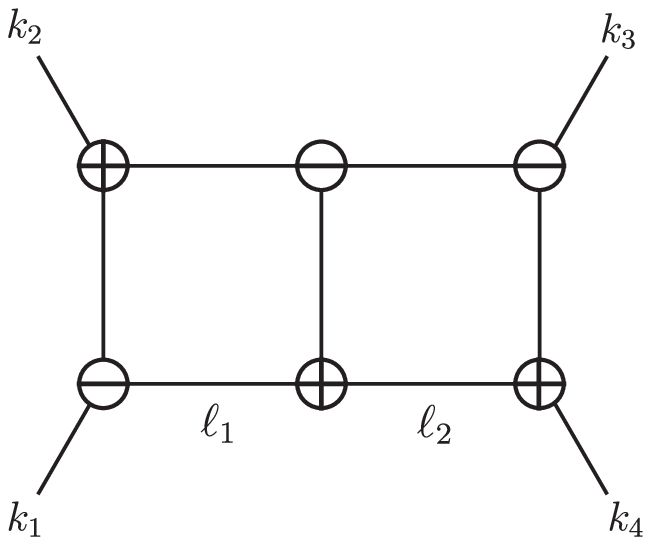} \end{center}
{\renewcommand{\arraystretch}{1.3}
\vspace{-5.1mm}
Solution $\Sol_3$, obtained by setting
\vspace{-5mm}
{\begin{eqnarray*}\begin{array}{ll}
\alpha_3 = 0\,, & \hspace{7mm} \beta_3 = 0\,, \nn\\[-5mm]
\alpha_4 = -\chi\,, & \hspace{7mm} \beta_4 = z\,.\nn
\end{array}
\end{eqnarray*}
}}
\end{minipage}
\begin{minipage}[b]{0.45\linewidth}
\begin{center} \includegraphics[scale=0.9]{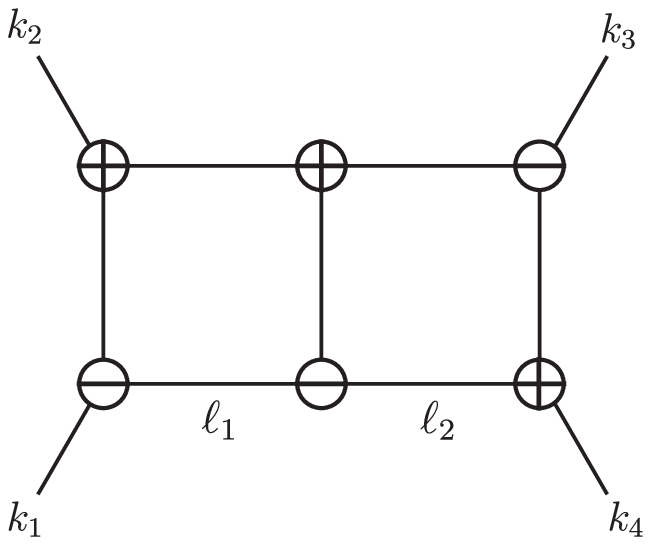} \end{center}
{\renewcommand{\arraystretch}{1.3}
\vspace{-5.1mm}
Solution $\Sol_4$, obtained by setting
\vspace{-5mm}
\begin{eqnarray*}\begin{array}{ll} 
\alpha_3 = 0\,, & \hspace{7mm} \beta_3 = 0\,,\nn \\[-5mm]
\alpha_4 = z\,, & \hspace{7mm} \beta_4 = -\chi\,.
\end{array}
\end{eqnarray*}}
\end{minipage}
%
%
\begin{minipage}[b]{0.45\linewidth}
\begin{center} \includegraphics[scale=0.9]{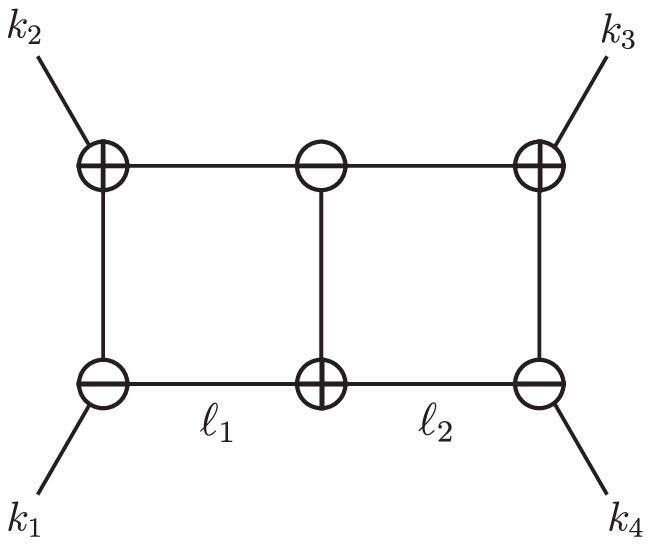} \end{center}
{\renewcommand{\arraystretch}{1.3}
\vspace{-5.1mm}
Solution $\Sol_5$, obtained by setting
\vspace{-5mm}
\begin{eqnarray*}\begin{array}{ll} 
\alpha_3 = 0\,, & \hspace{7mm} \beta_3 = -(\chi + 1) \frac{z + \chi}{z + \chi + 1}\,, \nn\\[-5mm]
\alpha_4 = z\,, & \hspace{7mm} \beta_4 = 0\,.
\end{array}
\end{eqnarray*}}
\end{minipage}
\begin{minipage}[b]{0.45\linewidth}
\begin{center} \includegraphics[scale=0.9]{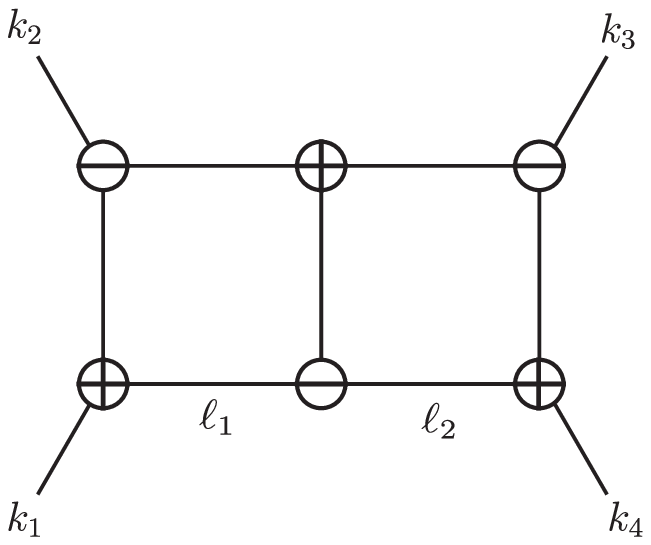} \end{center}
{\renewcommand{\arraystretch}{1.3}
\vspace{-5.1mm}
Solution $\Sol_6$, obtained by setting
\vspace{-5mm}
\begin{eqnarray*}\begin{array}{ll} 
\alpha_3 = z\,, & \hspace{7mm}\beta_3 = 0\,, \nn\\[-5mm]
\alpha_4 = 0\,, & \hspace{7mm} 
  \beta_4 = -(\chi + 1) \frac{z +\chi}{z + \chi + 1}\,.
\end{array}
\end{eqnarray*}}
\end{minipage}
\caption{The six solutions to the heptacut equations for the two-loop
planar double box.}
\label{SixSolutions}
\end{figure}

\def\tlambda{\tilde\lambda}
The existence of six kinematic solutions can also be understood from
holomorphicity considerations of the cuts.  When we cut
all propagators, each of the six vertices in the double box
has three massless momenta attached.  We can write these
momenta in terms of spinors, $k^\mu = \lambda^\alpha
\sigma_{\alpha \dot{\alpha}}^\mu
\tlambda^{\dot{\alpha}}$.  Momentum conservation
at each vertex~\cite{WittenTwistorString} then implies that either,
\begin{itemize}
\item[1)] the holomorphic spinors $\lambda$ of the momenta are 
collinear (proportional),
$\lambda_a \propto \lambda_b \propto \lambda_c$.  We will
depict such a vertex using a circled plus ($\oplus$).  Such
a vertex would allow only an $\overline{\mathrm{MHV}}$ tree amplitude
to be attached (of course the holomorphicity properties of the cut
are independent of any tree amplitude).

\item[2)] the antiholomorphic spinors $\tlambda$ of the momenta are
collinear, $\tlambda_a \propto \tlambda_b
\propto \tlambda_c$.  We will depict such a vertex using
a circled minus ($\ominus$). 
Such
a vertex would allow only an ${\mathrm{MHV}}$ tree amplitude
to be attached.
\end{itemize}
For general kinematics, neither the external holomorphic spinors
$\lambda_j$ nor the external antiholomorphic spinors $\tlambda_j$
are collinear.  A configuration with an uninterrupted chain of
either $\oplus$ or $\ominus$ vertices connecting any two external
legs is thus disallowed.  There are exactly six ways of assigning
these two labelings to vertices avoiding such chains, hence six
solutions.
The six solutions
are shown diagrammatically in \fig{SixSolutions}.
(The labeling of holomorphically-collinear vertices as $\oplus$,
and of antiholomorphically-collinear ones as $\ominus$ is not
uniform in the literature.)

In evaluating the contour integrals represented by the delta functions
in \eqn{HeptacutDoubleBox}, we encounter two Jacobians: one from
changing variables from the components of $\ell_j$ to the $\alpha_i$
and $\beta_i$; and one from actually performing the contour integrals
in the latter variables.  It is the latter Jacobian that is important
for our purposes.  The former Jacobian is equal to $J_\alpha J_\beta$,
where
\begin{equation}
J_\alpha = \det_{\mu,i} \frac{\partial \ell_1^\mu}{\partial \alpha_i} =
  -\frac{i s_{12}^2}{4\chi (\chi+1)} \: , \hspace{7mm} 
J_\beta = \det_{\mu,i}
\frac{\partial \ell_2^\mu}{\partial \beta_i} = 
  -\frac{i s_{12}^2}{4\chi
(\chi+1)}\,. \label{eq:Jacobians_from_l1l2_to_alpha_beta}
\end{equation}

To evaluate the latter Jacobian, we may note that three of
the delta functions (or equivalently three of the contour integrals)
involve only $\alpha$ variables, and three involve only $\beta$
variables.  We can thus split up the problem into three steps:
computing the Jacobian associated with $\ell_1$, that is with the
$\alpha$ variables alone; computing the Jacobian associated
with $\ell_2$, that is with the $\beta$ variables alone; and
finally, computing the Jacobian associated with the middle
propagator, involving both $\ell_1$ and $\ell_2$.  

For each of the six solutions, we must compute the Jacobian
independently.  As an example, consider the second solution $\Sol_2$.
The first
Jacobian arises from considering the integral,
\begin{eqnarray}
&&\int d\alpha_1 d\alpha_2 d\alpha_4 \hspace{0.5mm}
\delta \hspace{-0.5mm} \left[ s_{12} \left( \alpha_1 \alpha_2 +
\textstyle{\frac{\alpha_3 \alpha_4}{\chi (\chi + 1)}} \right)
\right] \delta \hspace{-0.5mm} \left[ s_{12} \left( (\alpha_1 -1)
\alpha_2 + \textstyle{\frac{\alpha_3 \alpha_4}{\chi
(\chi + 1)}} \right) \right] \nonumber \\
&\phantom{=}& \hspace{30mm} \times \hspace{0.5mm} \delta
\hspace{-1mm} \left[ s_{12} \left( (\alpha_1 - 1)(\alpha_2 - 1) +
\textstyle{\frac{\alpha_3 \alpha_4}{\chi (\chi + 1)}} \right)
\right]\,,
\end{eqnarray}
associated with the $\ell_1$ loop.  Define
\begin{equation}
\left( \hspace{-1mm} \begin{array}{c} g_1(\alpha_1,\alpha_2,\alpha_4) 
\\ g_2(\alpha_1,\alpha_2,\alpha_4) \\
g_3(\alpha_1,\alpha_2,\alpha_4) \end{array} \hspace{-1mm} \right)
\hspace{1.5mm}=\hspace{1.5mm} \left(
\begin{array}{c} s_{12} \left( \alpha_1 \alpha_2 +
\textstyle{\frac{\alpha_3
\alpha_4}{\chi (\chi + 1)}} \right) \\
s_{12} \left( (\alpha_1 - 1) \alpha_2 + \textstyle{\frac{\alpha_3
\alpha_4}{\chi (\chi + 1)}} \right) \\
s_{12} \left( (\alpha_1 - 1) (\alpha_2 - 1) +
\textstyle{\frac{\alpha_3 \alpha_4}{\chi (\chi + 1)}} \right)
\end{array} \right) \, .
\end{equation}
The first Jacobian is then,
\begin{equation}
J_1 = \det_{i,j} \frac{\partial g_i}{\partial \alpha_j}
\hspace{1.5mm}=\hspace{1.5mm} s_{12}^3 \det\left(
\begin{array}{ccc}
\alpha_2 & \alpha_1 & \frac{\alpha_3}{\chi (\chi + 1)} \\
\alpha_2 & \alpha_1 -1 & \frac{\alpha_3}{\chi (\chi + 1)} \\
\alpha_2 -1 & \alpha_1 -1 & \frac{\alpha_3}{\chi (\chi + 1)}
\end{array} \right)
 \hspace{1.5mm}=\hspace{1.5mm} -\frac{s_{12}^3}{\chi (\chi + 1)} \alpha_3
\end{equation}
(As explained in \Sect{ReviewSection}, the Jacobians will
appear in the denominator as determinants
rather than as absolute values of determinants.)
Similarly, the second Jacobian arises from considering the integral,
\begin{eqnarray}
&& \int d\beta_1 d\beta_2 d\beta_4 \hspace{0.5mm} \delta
\hspace{-0.5mm} \left[ s_{12} \left( \beta_1 \beta_2 +
\textstyle{\frac{\beta_3 \beta_4}{\chi (\chi + 1)}} \right)
\right] \delta \hspace{-0.5mm} \left[ s_{12} \left( \beta_1
(\beta_2 -1) + \textstyle{\frac{\beta_3 \beta_4}{\chi
(\chi + 1)}} \right) \right] \nonumber \\
&\phantom{=}& \hspace{30mm} \times \hspace{0.5mm} \delta
\hspace{-1mm} \left[ s_{12} \left( (\beta_1 - 1)(\beta_2 - 1) +
\textstyle{\frac{\beta_3 \beta_4}{\chi (\chi + 1)}} \right)
\right]\,,
\end{eqnarray}
associated with the $\ell_2$ loop.  Define
\begin{equation}
 \left( \hspace{-1mm} \begin{array}{c} 
h_1(\beta_1,\beta_2,\beta_4) \\ 
h_2(\beta_1,\beta_2,\beta_4) \\
h_3(\beta_1,\beta_2,\beta_4) \end{array} \hspace{-1mm} \right)
\hspace{1.5mm}=\hspace{1.5mm} \left(
\begin{array}{c} s_{12} \left( \beta_1 \beta_2 +
\textstyle{\frac{\beta_3
\beta_4}{\chi (\chi + 1)}} \right) \\
s_{12} \left( \beta_1 (\beta_2 - 1) + \textstyle{\frac{\beta_3
\beta_4}{\chi (\chi + 1)}} \right) \\
s_{12} \left( (\beta_1 - 1) (\beta_2 - 1) +
\textstyle{\frac{\beta_3 \beta_4}{\chi (\chi + 1)}} \right)
\end{array} \right) \, .
\end{equation}
The second Jacobian is then,
\begin{equation}
J_2 = \det_{i,j} \frac{\partial h_i}{\partial \beta_j}
\hspace{1.5mm}=\hspace{1.5mm} s_{12}^3 \det\left(
\begin{array}{ccc}
\beta_2 & \beta_1 & \frac{\beta_3}{\chi (\chi + 1)} \\
\beta_2 - 1 & \beta_1 & \frac{\beta_3}{\chi (\chi + 1)} \\
\beta_2 -1 & \beta_1 -1 & \frac{\beta_3}{\chi (\chi + 1)}
\end{array} \right) \hspace{1.5mm}=\hspace{1.5mm} \frac{s_{12}^3}{\chi (\chi + 1)} \beta_3\,.
\end{equation}
The remaining integration we must consider is over $\alpha_3$ and
$\beta_3$,
\begin{eqnarray}
&& \frac{1}{2} \int d \alpha_3 d\beta_3 \frac{J_\alpha J_\beta}{J_1 J_2}
 \hspace{0.8mm} \delta \hspace{-0.5mm}
    \left[\frac{s_{12}}{2\chi} (\alpha_3 + \chi)(\beta_3 +\chi) \right] =
  \nn\\ && \hskip 5mm
 \frac{1}{32 s_{12}^2} \int \frac{d \alpha_3 d\beta_3} {\alpha_3\beta_3}
 \hspace{0.8mm} \delta \hspace{-0.5mm}
    \left[\frac{s_{12}}{2\chi} (\alpha_3 + \chi)(\beta_3 +\chi) \right] 
 \label{eq:contribution_to_Jac_from_kin_sol_1_eq_2}\,,
\end{eqnarray}
which leaves a remaining contour integration over $z$ ({\it i.e.} $\alpha_3$),
 along with the overall
inverse Jacobian,
\begin{equation}
J^{-1}(z) = -\frac1{16 s_{12}^3\,z (z+\chi)}\,.
\end{equation}
The computation for the other five solutions is similar; it turns out
that we obtain the same overall Jacobian for all solutions.
The contour for the $z$ integration remains to be chosen; for this
solution, there are two possible non-trivial contours, one encircling
$z=0$, and the other, encircling $z=-\chi$.  (We set aside a 
possible non-trivial contour
encircling $z=\infty$, as its contribution when integrating an
arbitrary multiplying function $f(z)$ sums to zero when combined with
the contributions of these two contours.)  The pole 
at $z=-\chi$ is the
eighth pole in the octacut of ref.~\cite{Octacut}.  In addition, for
solutions $\Sol_{5,6}$, the denominator of $\beta_{3,4}$ 
(\eqns{Solution5}{Solution6}) can give rise
to additional poles at $z=-\chi-1$ in tensor integrals.
(As noted in \Sect{ReviewSection},
in a slight abuse of language, we refer
to integrals with no free indices, but numerator
powers of the loop momenta
contracted into external vectors, as ``tensor integrals''.)

  Collecting
the information above, we have the following contours we can
utilize in seeking equations for integral coefficients,
\begin{eqnarray}
T_{1,1}=&& T_0\times C_{\alpha_3}(-\chi)\times
  C_{\alpha_4}(0)\times C_{\beta_3=z}(0)\times C_{\beta_4}(0)\,,\nn\\
T_{1,2}=&& T_0\times C_{\alpha_3}(-\chi)\times
  C_{\alpha_4}(0)\times C_{\beta_3=z}(-\chi)\times C_{\beta_4}(0)\,,\nn\\
T_{2,1}=&& T_0\times C_{\alpha_3=z}(0)\times
  C_{\alpha_4}(0)\times C_{\beta_3}(-\chi)\times C_{\beta_4}(0)\,,\nn\\
T_{2,2}=&& T_0\times C_{\alpha_3=z}(-\chi)\times
  C_{\alpha_4}(0)\times C_{\beta_3}(-\chi)\times C_{\beta_4}(0)\,,\nn\\
T_{3,1} =&& T_0\times C_{\alpha_3}(0)\times 
  C_{\alpha_4}(-\chi)\times C_{\beta_3}(0)\times C_{\beta_4=z}(0)\,,\nn\\
T_{3,2} =&& T_0\times C_{\alpha_3}(0)\times 
  C_{\alpha_4}(-\chi)\times C_{\beta_3}(0)\times C_{\beta_4=z}(-\chi)\,,\nn\\
T_{4,1}=&& T_0\times C_{\alpha_3}(0)\times
  C_{\alpha_4=z}(0)\times C_{\beta_3}(0)\times C_{\beta_4}(-\chi)\,,\nn\\
T_{4,2}=&& T_0\times C_{\alpha_3}(0)\times
  C_{\alpha_4=z}(-\chi)\times C_{\beta_3}(0)\times C_{\beta_4}(-\chi)\,,
\label{TwoLoopContours}\\
T_{5,1}=&& T_0\times C_{\alpha_3}(0)\times
  C_{\alpha_4=z}(0)
    \times C_{\beta_3}\left(-\frac{(1+\chi)(z+\chi)}{z+\chi+1}\right)
     \times C_{\beta_4}(0)\,,\nn\\
T_{5,2}=&& T_0\times C_{\alpha_3}(0)\times
  C_{\alpha_4=z}(-\chi)
    \times C_{\beta_3}\left(-\frac{(1+\chi)(z+\chi)}{z+\chi+1}\right)
     \times C_{\beta_4}(0)\,,\nn\\
T_{5,3}=&& T_0\times C_{\alpha_3}(0)\times
  C_{\alpha_4=z}(-\chi-1)
    \times C_{\beta_3}\left(-\frac{(1+\chi)(z+\chi)}{z+\chi+1}\right)
     \times C_{\beta_4}(0)\,,\nn\\
T_{6,1}=&& T_0\times C_{\alpha_3=z}(0)\times
  C_{\alpha_4}(0)
    \times C_{\beta_3}(0)
     \times C_{\beta_4}\left(-\frac{(1+\chi)(z+\chi)}{z+\chi+1}\right)\,,\nn\\
T_{6,2}=&& T_0\times C_{\alpha_3=z}(-\chi)\times
  C_{\alpha_4}(0)
    \times C_{\beta_3}(0)
     \times C_{\beta_4}\left(-\frac{(1+\chi)(z+\chi)}{z+\chi+1}\right)\,,\nn\\
T_{6,3}=&& T_0\times C_{\alpha_3=z}(-\chi-1)\times
  C_{\alpha_4}(0)
    \times C_{\beta_3}(0)
     \times C_{\beta_4}\left(-\frac{(1+\chi)(z+\chi)}{z+\chi+1}\right)\,,\nn
\end{eqnarray}
where each subscript denotes the variable in whose plane the circle lies,
and where
\begin{equation}
T_0 = C_{\alpha_1}(1) \times C_{\alpha_2}(0)
  \times C_{\beta_1}(0) \times C_{\beta_2}(1)\,,
\end{equation}
corresponding to the on-shell values in \eqn{AlphaBetaEquations}.
 We will call the complete
contours, including a choice of contour for $z$, the `augmented heptacut'.

Naively, we could deform the original contour of integration for the
double box~(\ref{eq:double_box_integral_def}), along the product
of real axes for all components of $\ell_1$ and $\ell_2$, to any
linear combination of contours in \eqn{TwoLoopContours} that we wish.
However, an arbitrary deformation will not preserve the vanishing
of total derivatives, analogs to $U_1$ given in \eqn{TotalDerivativeExample}.
In order to ensure that such objects vanish as they must, we impose
constraints on the contours.  We derive these in the next section.

\section{Constraint Equations for Contours}
\label{ConstraintSection}

Integral reductions are implicitly part of the simplifications
applied to a sum over Feynman diagrams in order to obtain the
basic equation at either one loop~(\ref{BasicEquation}), or at
two loops,
\begin{equation}
{\rm Amplitude} = \sum_{j\in \mathrm{Two\mbox{-}Loop\ Basis}} 
  {\rm coefficient}_j {\rm Integral}_j + 
{\rm Rational}\,.
\label{TwoLoopBasicEquation}
\end{equation}
The basis at two loops will contain integrals with up to eight 
propagators in the planar case~\cite{TwoLoopBasis}, 
though a specific complete and independent choice of integrals
for a general amplitude
has not yet been written down.  (The same restriction to eight propagators
or fewer presumably applies in the non-planar case as well, using
arguments along the same lines as given in ref.~\cite{TwoLoopBasis}.)

As we saw in \Sect{ReviewSection},
integral reductions at one loop involve only rewriting dot products
of the loop momentum in terms of linear combinations
of propagators and external invariants, along with the use of
Lorentz invariance and parity to eliminate some integrals.  For the
box integral, in particular, 
the only non-trivial constraint arises
from the use of parity, which requires that
\begin{equation}
\int {d^D\ell\over (2\pi)^D}\; 
  {\varepsilon(\ell,k_1,k_2,k_4)
   \over \ell^2 (\ell-k_1)^2 (\ell-k_1-k_2)^2(\ell+k_4)^2} = 0\,.
\end{equation}
This constraint must be respected by the unitarity procedure; otherwise,
applying a cut to the original integral and to the integral after
reduction would yield different, and hence inconsistent, answers.
At one loop, it gives rise to one constraint equation, which fixes
the relative normalization of the contours encircling the two solutions
to the on-shell equations.

Similar constraints arise at two loops, though we have a greater
variety of Levi-Civita symbols to consider.
Denoting the insertion of the function $f(\ell_1,\ell_2)$ in the
numerator of the double box by
\begin{equation}
\DBox[f(\ell_1,\ell_2)] = 
\int {d^{D} \ell_1\over (2\pi)^D} \hspace{0.5mm} {d^{D} \ell_2\over(2\pi)^D}
\hspace{0.7mm} \frac{f(\ell_1,\ell_2)}{\ell_1^2(\ell_1 - k_1)^2
(\ell_1 - K_{12})^2(\ell_1 + \ell_2)^2\ell_2^2
(\ell_2 - k_4)^2(\ell_2 - K_{34})^2} \, ,
\label{DoubleBoxWithInsertion}
\end{equation}
we must require that the vanishing of the following integrals,
\begin{eqnarray}
&& \DBox[\varepsilon(\ell_1,k_2,k_3,k_4)]\,,\quad
\DBox[\varepsilon(\ell_2,k_2,k_3,k_4)]\,,\quad
\DBox[\varepsilon(\ell_1,\ell_2,k_1,k_2)]\,,\quad\nn\\
&&
\DBox[\varepsilon(\ell_1,\ell_2,k_1,k_3)]\,,\quad
\mathrm{and}\quad\DBox[\varepsilon(\ell_1,\ell_2,k_2,k_3)]\,,\quad
\label{LeviCivitaConstraints}
\end{eqnarray}
continues to hold for integration over our chosen linear combination
of contours.  This is the complete set of Levi-Civita symbols that
arises during integral reduction, after using momentum conservation.

At two loops, additional reductions are required in order to arrive at
a linearly-independent set of basis integrals.  These are usually
obtained through IBP relations~\cite{IBP,Laporta,AIR,FIRE,Reduze};
that is, they correspond to adding a total derivative to the original
integrand.  Each such total derivative, or equivalently each
non-trivial reduction identity, gives rise to a constraint requiring
that the unitarity procedure give vanishing coefficients for the
additional terms; or equivalently, that the unitarity procedure
respect the reduction equations.  This is not automatically true
contour-by-contour, and hence gives rise to non-trivial constraints on
the choice of contours, and the weighting of different solutions.

In two-loop four-point amplitudes, we can express all dot products
of loop momenta with external vectors in terms of eight dot products:
$\ell_j\cdot k_1$, $\ell_j\cdot k_2$, $\ell_j\cdot k_4$, and
$\ell_j\cdot v$, where $v^\mu=\varepsilon(\mu,k_1,k_2,k_4)$.
Just as at one loop,
odd powers of $v$ will give rise to vanishing integrals,
as expressed in the Levi-Civita constraints discussed above. Even
powers can again be re-expressed in terms of the other dot products
(up to terms involving the $(-2\eps)$-dimensional components of
the loop momentum).  All integrals can then be rewritten in terms
of the six dot products of the loop momenta with
the external momenta.

Of these six dot products, three of them --- $\ell_1\cdot k_1$,
$\ell_1\cdot k_2$, $\ell_2\cdot k_4$ ---
can be rewritten as linear combinations of the propagator denominators
and external invariants.  One additional dot product of $\ell_2$ --- say
$\ell_2\cdot k_2$ --- can be rewritten in terms of the remaining two
($\ell_1\cdot k_4$ and $\ell_2\cdot k_1$), propagator denominators, and
external invariants.  The remaining two dot products are called
irreducible.  At a first stage, then, before using IBP identities,
we can reduce an arbitrary double-box integral appearing in a gauge-theory
amplitude to a linear combination of the 22 different integrals that
can arise with powers of the two irreducible numerators.

We have the following naively-irreducible integrals,
\begin{eqnarray}
&\DBox[1],
\DBox[ \ell_{2}\cdot k_{1}],\,\,
\DBox[ (\ell_{2}\cdot k_{1})^{2}],\,\,
\DBox[( \ell_{2}\cdot k_{1})^{3}],\,\,
\DBox[ \ell_{1}\cdot k_{4}],
\nn\\&
\DBox[( \ell_{2}\cdot k_{1})\,\,( \ell_{1}\cdot k_{4})],\,\,
\DBox[( \ell_{2}\cdot k_{1})^{2}\,\,( \ell_{1}\cdot k_{4})],\,\,
\DBox[( \ell_{2}\cdot k_{1})^{3}\,\,( \ell_{1}\cdot k_{4})],\,\,
\nn\\&
\DBox[( \ell_{1}\cdot k_{4})^{2}],\,\,
\DBox[( \ell_{2}\cdot k_{1})\,\,( \ell_{1}\cdot k_{4})^{2}],\,\,
\DBox[( \ell_{2}\cdot k_{1})^{2}\,\,( \ell_{1}\cdot k_{4})^{2}],
\nn\\&
\DBox[( \ell_{2}\cdot k_{1})^{3}\,\,( \ell_{1}\cdot k_{4})^{2}],\,\,
\DBox[( \ell_{1}\cdot k_{4})^{3}],\,\,
\DBox[( \ell_{2}\cdot k_{1})\,\,( \ell_{1}\cdot k_{4})^{3}],
\label{DoubleBoxTensorIntegrals} \\&
\DBox[( \ell_{2}\cdot k_{1})^{2}\,\,( \ell_{1}\cdot k_{4})^{3}],\,\,
\DBox[( \ell_{2}\cdot k_{1})^{3}\,\,( \ell_{1}\cdot k_{4})^{3}],\,\,
\DBox[( \ell_{1}\cdot k_{4})^{4}],
\nn\\&
\DBox[( \ell_{2}\cdot k_{1})\,\,( \ell_{1}\cdot k_{4})^{4}],\,\,
\DBox[( \ell_{2}\cdot k_{1})^{2}\,\,( \ell_{1}\cdot k_{4})^{4}],\,\,
\DBox[( \ell_{2}\cdot k_{1})^{4}],
\nn\\&
\DBox[( \ell_{2}\cdot k_{1})^{4}\,\,( \ell_{1}\cdot k_{4})],\,\,
\DBox[( \ell_{2}\cdot k_{1})^{4}( \ell_{1}\cdot k_{4})^{2}]\,.
\nn
\end{eqnarray}
In the massless case, it turns out that there are 20 IBP relations
between these integrals, which allow further
reductions.  These reductions allow us to pick certain pairs,
for example,
\begin{equation}
\DBox[1]\mathrm{\ \ and\ \ }
\DBox[\ell_{1}\cdot k_{4}]\,,
\label{BasisIntegrals}
\end{equation}
as master integrals for the set in \eqn{DoubleBoxTensorIntegrals}, and
thus also as basis integrals for an amplitude.

The remaining integrals are given in terms of these two by linear
equations, for example
\begin{eqnarray}
\DBox[\ell_2 \cdot k_1] \hspace{-2mm} &=& \hspace{-2mm} \DBox[\ell_1 \cdot k_4]\,, \nn\\
\DBox[(\ell_1 \cdot k_4)(\ell_2 \cdot k_1)] \hspace{-2mm} &=& \hspace{-2mm}
\frac{1}{8} \chi s_{12}^2 \DBox[1] - \frac{3}{4} s_{12} \DBox[\ell_1 \cdot k_4] +\cdots\,, \nn\\
\DBox[(\ell_2 \cdot k_1)^2] \hspace{-2mm} &=& \hspace{-2mm}
-\frac{\epsilon \chi s_{12}^2}{4(1 - 2 \epsilon)} \DBox[1]
+\frac{(\chi + 3\epsilon)s_{12}}{2(1-2\epsilon)} \DBox[\ell_1 \cdot k_4] +\cdots\,, \nn\\
\DBox[(\ell_1 \cdot k_4)^2] \hspace{-2mm} &=& \hspace{-2mm}
-\frac{\epsilon \chi s_{12}^2}{4(1 - 2 \epsilon)} \DBox[1]
+ \frac{(\chi + 3\epsilon)s_{12}}{2(1-2\epsilon)} \DBox[\ell_1 \cdot k_4] +\cdots\,, \nn\\
\DBox[(\ell_1 \cdot k_4)(\ell_2 \cdot k_1)^2] \hspace{-2mm} &=& -\frac{(1 -
3\epsilon) \chi s_{12}^3}{16(1 - 2\epsilon)} \DBox[1] +
\frac{(3-9\epsilon - 2\epsilon\chi)s_{12}^2}
{8(1-2\epsilon)} \DBox[\ell_1 \cdot k_4] +\cdots\,, 
\label{eq:IBP-relation_8}\\
\DBox[(\ell_1 \cdot k_4)^2(\ell_2 \cdot k_1)] \hspace{-2mm} &=& -\frac{(1 -
3\epsilon) \chi s_{12}^3}{16(1 - 2\epsilon)} \DBox[1] +
\frac{(3-9\epsilon - 2\epsilon\chi)s_{12}^2}
{8(1-2\epsilon)} \DBox[\ell_1 \cdot k_4] +\cdots\,, \nn\\
\DBox[(\ell_2 \cdot k_1)^3] \hspace{-2mm} &=& \frac{\epsilon \chi
(1-2\chi - 3\epsilon) s_{12}^3}{16 (1-\epsilon)(1 - 2\epsilon)}
\DBox[1] \nn\\
&&\hskip 15mm
+ \frac{\big(2\chi^2 - 3\epsilon (1-2\chi) + \epsilon^2 (9 +
2\chi) \big)s_{12}^2}{8 (1-\epsilon)(1-2\epsilon)} \DBox[\ell_1 \cdot
k_4] +\cdots\,, \nn\\
\DBox[(\ell_1 \cdot k_4)^3] \hspace{-2mm} &=& \frac{\epsilon \chi
(1-2\chi - 3\epsilon) s_{12}^3}{16 (1-\epsilon)(1 - 2\epsilon)}
\DBox[1] \nn\\
&&\hskip 15mm
+ \frac{\big(2\chi^2 - 3\epsilon (1-2\chi) + \epsilon^2 (9 +
2\chi) \big)s_{12}^2}{8 (1-\epsilon)(1-2\epsilon)} \DBox[\ell_1 \cdot
k_4] +\cdots\,, \nn
\end{eqnarray}
where the ellipses denote additional integrals with fewer propagators.
We must require that these equations (and the other 12 we do not
display explicitly) are preserved by the choice of contours.  The
contour integrals which implement the augmented heptacut will yield
vanishing results for the integrals with fewer propagators, so they do
not enter the constraint equations. As we are considering
only four-dimensional cuts, the augmented heptacuts are effectively
four-dimensional.

In order to find the explicit form of the constraint equations, 
denote the weight of contour $T_{j,r}$ by $a_{r,j}$,
\begin{eqnarray}
a_{1,j} &\longrightarrow& \mathrm{encircling} \hspace{2mm} z=0
\hspace{2mm} \mbox{for solution} \hspace{1.2mm} \Sol_j\,,\nn\\
a_{2,j} &\longrightarrow& \mathrm{encircling} \hspace{2mm} z=-\chi
\hspace{2mm} \mbox{for solution} \hspace{1.2mm} \Sol_j\,,
\label{eq:notation_for_heptacut_Jac_poles_3}\\
a_{3,j} &\longrightarrow& \mathrm{encircling} \hspace{2mm} z=-\chi-1
\hspace{2mm} \mbox{for solution} \hspace{1.2mm} \Sol_j\,.\nn
\end{eqnarray}

For a numerator insertion of $f(\ell_1,\ell_2)$ in the numerator
of the double box, the augmented heptacut is then,
\begin{eqnarray}
&&\sum_{j=1}^4 \sum_{r=1}^2 a_{r,j} \oint_{T_{j,r}} 
d^4\alpha_i d^4\beta_i\;f(\ell_1,\ell_2)\times {\rm Propagators}(\ell_1,\ell_2)
\bigg|_{\rm param}
\nn\\
&&+\sum_{j=5}^6 \sum_{r=1}^3 a_{r,j} \oint_{T_{j,r}} 
d^4\alpha_i d^4\beta_i\;f(\ell_1,\ell_2)\times {\rm Propagators}(\ell_1,\ell_2)
\bigg|_{\rm param}
\,,
\label{GeneralAugementHeptacut}
\end{eqnarray}
where the notation $|_{\rm param}$
indicates that we use the parametrization of $\ell_1$ and $\ell_2$ 
given in \eqn{TwoLoopParametrization}.
The signs in front of each coefficient $a_{r,j}$ in the result
will depend on the orientation chosen for the
corresponding contour; but this sign will drop out of final formul\ae{} for
integral coefficients so long as this orientation is chosen
consistently throughout the calculation.

We can write down a compact expression for the 
augmented heptacut of the general tensor integral,
\begin{eqnarray}
\DBox[(\ell_1 \cdot k_4)^m (\ell_2 \cdot k_1)^n] \hspace{-1mm} &=&
   \hspace{-1mm} -\frac{1}{128} \left[ \hspace{0.4mm}
   \delta_{m,0} \left(
\frac{s_{12}}{2} \right)^{n-3} 
   \oint_{\Gamma_1} \frac{dz}{z} (z+\chi)^{n-1} \right.\nn \\
&\phantom{=}& 
 +\delta_{n,0}
\left( \frac{s_{12}}{2} \right)^{m-3} 
   \oint_{\Gamma_2} \frac{dz}{z} (z+\chi)^{m-1} \nn \\
&\phantom{=}& + \hspace{1mm} \delta_{m,0}
\left( \frac{s_{12}}{2} \right)^{n-3} 
   \oint_{\Gamma_3} \frac{dz}{z} (z+\chi)^{n-1}\nn\\
&\phantom{=}& + \hspace{1mm} \delta_{n,0} \left(
\frac{s_{12}}{2} \right)^{m-3} 
   \oint_{\Gamma_4} \frac{dz}{z} (z+\chi)^{m-1} \nn \\
&\phantom{=}& + \left( \frac{s_{12}}{2}
\right)^{m+n-3} \oint_{\Gamma_5} \frac{dz}{z}
(z+\chi)^{m-1} \left( -\frac{z}{z+\chi +1} \right)^n \nn \\
&\phantom{=}&  
 \left. + \left( \frac{s_{12}}{2}
\right)^{m+n-3} \oint_{\Gamma_6} \frac{dz}{z}
(z+\chi)^{m-1} \left( - \frac{z}{z+\chi +1} \right)^n  \right]  \,,
\label{eq:heptacut_tensor_integral}
\end{eqnarray}
where $\Gamma_j$ denotes the $z$ component of $\sum_r a_{r,j}T_{j,r}$,
and where in our notation,
the contour integral implicitly includes a factor of
$1/(2\pi i)$, as noted in \Sect{ReviewSection}.

We can evaluate this expression using the contours as weighted
in \eqn{GeneralAugementHeptacut}; we find,
\begin{eqnarray}
\DBox[1] \hspace{-1.5mm}&=&\hspace{-1.5mm} 
-\frac{1}{16 \chi s_{12}^3} \sum_{j=1}^{6} (a_{1,j} - a_{2,j})\,, \label{eq:cross_power_relation_1}\\
\DBox[(\ell_1 \cdot k_4)^m] \hspace{-1.5mm}&=&\hspace{-1.5mm} 
-\frac{1}{32s_{12}^2} \left( \frac{\chi s_{12}}{2} \right)^{m-1}
\sum_{j\neq 1,3} a_{1,j}\,, \label{eq:cross_power_relation_2}\\
\DBox[(\ell_2 \cdot k_1)^n] \hspace{-1.5mm}&=&\hspace{-1.5mm} 
-\frac{1}{32s_{12}^2} \left( \frac{\chi s_{12}}{2} \right)^{n-1}
(-a_{2,6} + a_{3,6} - a_{2,5} + a_{3,5} + a_{1,1} + a_{1,3})\,, \label{eq:cross_power_relation_3}\\
\DBox[(\ell_1 \cdot k_4)^m (\ell_2 \cdot k_1)^n]
\hspace{-1.5mm}&=&\hspace{-1.5mm} \frac{1}{64 s_{12}} \left(
-\frac{s_{12}}{2} \right)^{m+n-2}
\Big[\Theta\left(\min(m,n)-\textstyle{\frac{5}{2}}\right) (\chi + 1)(\chi +2)
 \nonumber \\
&\phantom{=}& \hspace{4mm} + \hspace{0.7mm}
\Theta\left(\min(m,n)-\textstyle{\frac{3}{2}}\right) (m+n-3) (\chi + 1)
+\Theta\left(\min(m,n)-\textstyle{\frac{1}{2}}\right)\Big]\nn\\
&&\hskip 5mm\times  (a_{3,6}+
a_{3,5})\,, \phantom{aaaaaa} \label{eq:cross_power_relation_4}
\end{eqnarray}
where $m,n\geq 1$ and the last result is valid only
for $0\leq m+n \leq 6$ and $0\leq m,n \leq 4$ 
(corresponding to the numerator
insertions allowed in gauge theory in $D=4-2\epsilon$ dimensions).

With these expressions, we now turn to the constraint equations.  Let us
begin with the equations arising from the insertion of Levi-Civita
tensors~(\ref{LeviCivitaConstraints}).  Start with 
$\varepsilon(\ell_1,k_2,k_3,k_4)$,
\begin{eqnarray}
0 &=&  \DBox[\varepsilon(\ell_1, k_2, k_3, k_4)] 
\hskip5mm\Longrightarrow\nn \\
0&=& \hspace{-2mm} -\frac{1}{16 s_{12}^3} \left( 
\oint_{\Gamma_1} \!\frac{dz}{z}
\frac{\varepsilon\!\left(k_1^\mu - \frac{s_{12}}{2} \frac{
\sand1.{\gamma^\mu}.2}{\sand1.4.2} \chi, k_2, k_3, k_4 \right)}{z+\chi} 
  + \oint_{\Gamma_2} \!\frac{dz}{z} \frac{\varepsilon\!\left(k_1^\mu +\frac{s_{12}}{2}
\frac{\sand1.{\gamma^\mu}.2}{\sand1.{4}.2}z, k_2, k_3, k_4 \right)}{z+\chi} 
   \right. \nn \\
&\phantom{=}& \hspace{2mm} 
  + \oint_{\Gamma_3} \!\frac{dz}{z} \frac{\varepsilon\!\left(k_1^\mu -
\frac{s_{12}}{2} \frac{\sand2.{\gamma^\mu}.1}{\sand2.4.1} \chi, k_2, k_3, k_4 \right)}{z+\chi}
  +\oint_{\Gamma_4} \!\frac{dz}{z} \frac{\varepsilon\!\left(k_1^\mu +
\frac{s_{12}}{2} \frac{\sand2.{\gamma^\mu}.1}{\sand2.4.1}z, k_2, k_3, k_4 \right)}{z+\chi}  \nn\\
&\phantom{=}& \hspace{2mm} \left. 
 +  \oint_{\Gamma_5} \!\frac{dz}{z}
\frac{\varepsilon\!\left(k_1^\mu + \frac{s_{12}}{2} 
  \frac{\sand2.{\gamma^\mu}.1}{\sand2.4.1}z, k_2, k_3, k_4 \right)}{z+\chi}
 +\oint_{\Gamma_6} \!\frac{dz}{z} \frac{\varepsilon\!\left(k_1^\mu +
\frac{s_{12}}{2} \frac{\sand1.{\gamma^\mu}.2}{\sand1.4.2}z, k_2, k_3, k_4 \right)}{z+\chi}\right)\,. \nn\\
\end{eqnarray}
Evaluating this expression on the augmented 
heptacut~(\ref{GeneralAugementHeptacut}), we obtain,
\begin{eqnarray}
&& \hspace{-2mm} \frac{1}{32 s_{12}^2} \left[ (a_{2,2} + a_{2,6}
- a_{1,1} + a_{2,1}) \hspace{0.7mm} \varepsilon
\hspace{-0.7mm}
 \left(\frac{\sand1.{\gamma^\mu}.2}{\sand1.4.2}, k_2, k_3, k_4 \right) \right. \nn \\
&\phantom{=}& \hspace{10mm} \left. + \hspace{1mm} (a_{2,5} +
a_{2,4} - a_{1,3} + a_{2,3}) \hspace{0.7mm} \varepsilon
\hspace{-0.7mm} \left(\frac{\sand2.{\gamma^\mu}.1}{\sand2.4.1}, k_2, k_3,
k_4 \right) \right] \label{Levi-CivitaConstraint1}\\
&=& \hspace{-2mm} \frac{1}{32 s_{12}^2} \big( a_{2,2} + a_{2,6} -
a_{2,5} - a_{2,4} - a_{1,1} + a_{2,1} + a_{1,3} - a_{2,3} \big)
\varepsilon \hspace{-0.7mm}\left(
\frac{\sand1.{\gamma^\mu}.2}{\sand1.4.2}, k_2, k_3, k_4 \right)\,,\nn
\end{eqnarray}
where the last line follows from the fact that the two Levi-Civita
symbols appearing on the first line are equal but opposite in value.

Similarly, from the insertion of $\varepsilon(\ell_2, k_2, k_3, k_4)$
one finds
\begin{eqnarray}
0  &=&  \DBox[\varepsilon(\ell_2, k_2, k_3, k_4)] 
 \hskip 5mm\Longrightarrow\nn \\
0&=& \hspace{-1mm} -\frac{1}{32 s_{12}^2} \big( -a_{1,2} + a_{2,2} +
a_{1,6} - a_{3,6} - a_{1,5} + a_{3,5} + a_{1,4} - a_{2,4} +
a_{2,1} - a_{2,3} \big) \nn \\
&\phantom{=}& \hspace{15mm} \times \hspace{0.5mm} \varepsilon
\hspace{-0.7mm}\left(\frac{\sand3.{\gamma^\mu}.4}{\sand3.1.4}, k_2, k_3, k_4 \right)\,, 
\label{Levi-CivitaConstraint2}
\end{eqnarray}
and from the insertion of $\varepsilon(\ell_1, \ell_2, k_i, k_j)$
with $(i,j) \in \{ (1,2), (1,3), (2,3)\}$ one finds
\begin{eqnarray}
0 &=& \DBox[\varepsilon(\ell_1, \ell_2, k_1, k_2)] 
\hskip 5mm\Longrightarrow\nn\\
0 &=& -\frac{1}{32 s_{12}^2} \big( a_{2,6}
- a_{3,6} - a_{2,5} + a_{3,5} - a_{1,1} + a_{1,3} \big)
\hspace{0.7mm} \varepsilon \hspace{-0.7mm}\left(\frac{
\sand1.{\gamma^\mu}.2}{\sand1.4.2},
k_2, k_3, k_4 \right)\,,\nn\\
0 &=& \DBox[\varepsilon(\ell_1, \ell_2, k_1, k_3)] 
\hskip 5mm\Longrightarrow\nn\\
0 &=& \frac{1}{32 s_{12}^2}\big( a_{2,6} - a_{2,5} - a_{1,1} + a_{1,3}\big)
\varepsilon
\hspace{-0.7mm}\left(\frac{\sand1.{\gamma^\mu}.2}{\sand1.4.2}, k_2, k_3,
k_4 \right)\,,\label{Levi-CivitaConstraint3}\\
0 &=& \DBox[\varepsilon(\ell_1, \ell_2, k_2, k_3)] 
\hskip 5mm\Longrightarrow\nn\\
0 &=& -\frac{1}{32 s_{12}^2} \big( a_{1,2} - a_{1,6} + a_{2,6} + a_{1,5}
- a_{2,5} - a_{1,4} - a_{1,1} + a_{1,3} \big) \hspace{0.7mm}
\nn\\ &&\hskip 20mm\times
\varepsilon \hspace{-0.7mm}\left(\frac{\sand1.{\gamma^\mu}.2}{\sand1.4.2},
k_2, k_3, k_4 \right)\,.
\phantom{aaaaaa} \nn
\end{eqnarray}

These results combine to give the constraints,
\begin{eqnarray}
a_{2,2} + a_{2,6} - a_{2,5} - a_{2,4} - a_{1,1} + a_{2,1} +
a_{1,3} - a_{2,3} &=& 0\,, \nn\\
a_{1,2} - a_{2,2} - a_{1,6} + a_{3,6} + a_{1,5} - a_{3,5} -
a_{1,4} + a_{2,4} - a_{2,1} + a_{2,3} &=& 0\,, \nn\\
a_{2,6} - a_{3,6} - a_{2,5} + a_{3,5} - a_{1,1} + a_{1,3} &=& 0\,, \\
a_{2,6} - a_{2,5} - a_{1,1} + a_{1,3} &=& 0\,, \nn\\
a_{1,2} - a_{1,6} + a_{2,6} + a_{1,5} - a_{2,5} - a_{1,4} -
a_{1,1} + a_{1,3} &=& 0 \,,\nn
\end{eqnarray}
or equivalently,
\begin{eqnarray}
a_{1,2} - a_{1,6} + a_{1,5} - a_{1,4} &=& 0\,, \nn \\
a_{2,2} - a_{2,4} + a_{2,1} - a_{2,3} &=& 0\,, \nn \\
a_{2,6} - a_{2,5} - a_{1,1} + a_{1,3} &=& 0\,,  \label{Levi-CivitaConstraintFinal}\\
a_{3,6} - a_{3,5} &=& 0\,. \nn
\end{eqnarray}
This set has one equation less: not all the equations from
the Levi-Civita symbols are independent.
We see that these equations are solved by insisting that the contours
for complex-conjugate pairs of solutions ($\Sol_1\leftrightarrow \Sol_3$,
$\Sol_2\leftrightarrow\Sol_4$, and $\Sol_5\leftrightarrow\Sol_6$)
carry equal weights.  This nicely generalizes the one-loop constraint
on contours.  However, these are not the only possible solutions; solutions
which do not insist complex-conjugate pairs carry equal weight are also
possible.

We next impose the constraints following from the IBP reductions.
Evaluating both sides of equations~(\ref{eq:IBP-relation_8}) 
along with the remaining 
12 reduction equations not displayed above, and
setting $\eps=0$, we find two additional
constraint equations,
\begin{eqnarray}
a_{1,2} + a_{1,6} + a_{1,5} + a_{1,4} &=& -a_{2,6} + a_{3,6} -
a_{2,5} + a_{3,5} + a_{1,1} + a_{1,3}\,,  \nn\\
a_{3,6} + a_{3,5} &=& -\frac{1}{2} \sum_{j=1}^6 (a_{1,j} -
a_{2,j}) + \frac{3}{2} \sum_{j\neq1,3} a_{1,j} \, .
\label{eq:heptacut_contour_eq_6}
\end{eqnarray}
In principle, one might expect 18 additional equations from the
remaining reduction identities; but these all turn out to be
automatically satisfied on the solutions of this pair of
equations.

Beyond ensuring that all the reduction identities are valid, we ultimately
want to determine the coefficients of the two basis 
integrals~(\ref{BasisIntegrals}).  Because the system of equations
leaves many undetermined weights $a_{r,j}$, we have the freedom
to choose values which also kill one or the other of the basis integrals.
That is, we can choose contours for which one or the other of the 
basis integrals has vanishing augmented heptacut.  To project out
the second basis integral, $\DBox[\ell_1\cdot k_4]$, we should also
require that \eqn{eq:cross_power_relation_2} with $m=1$ vanish,
\begin{equation}
\sum_{j\neq 1,3} a_{1,j} = 0\,.
\end{equation}
To project out the first basis integral, $\DBox[1]$, we should require
that \eqn{eq:cross_power_relation_1} vanish,
\begin{equation}
\sum_{j=1}^{6} (a_{1,j} - a_{2,j}) = 0\,.
\end{equation}

\begin{figure}[!h]
\begin{minipage}[b]{0.45\linewidth}
\begin{center}
\includegraphics[angle=0, width=\textwidth]{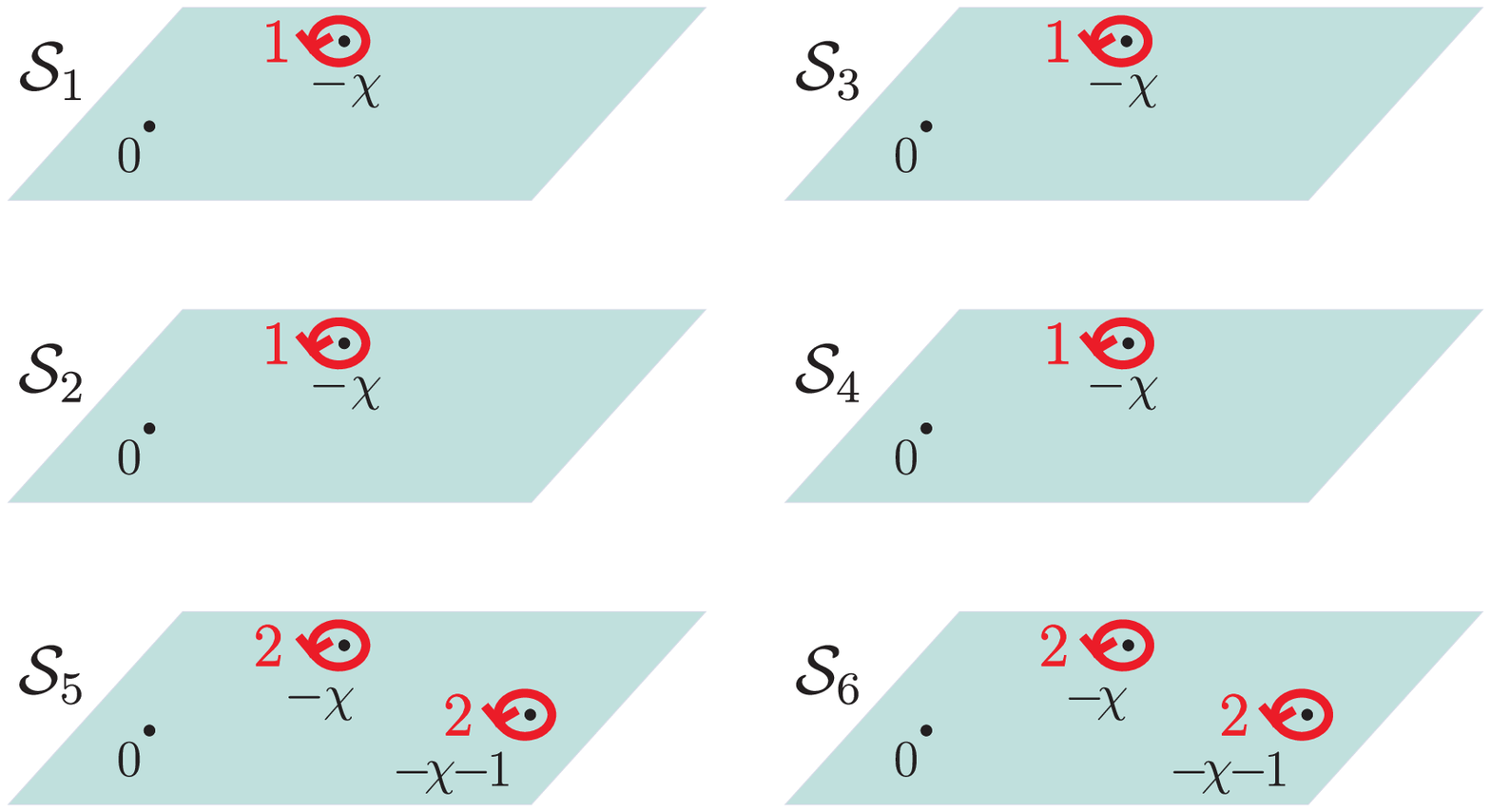}
\\{(a)}
\end{center}
\end{minipage}
\hskip 0.4truein
\begin{minipage}[b]{0.45\linewidth}
\begin{center}
\includegraphics[angle=0, width=\textwidth]{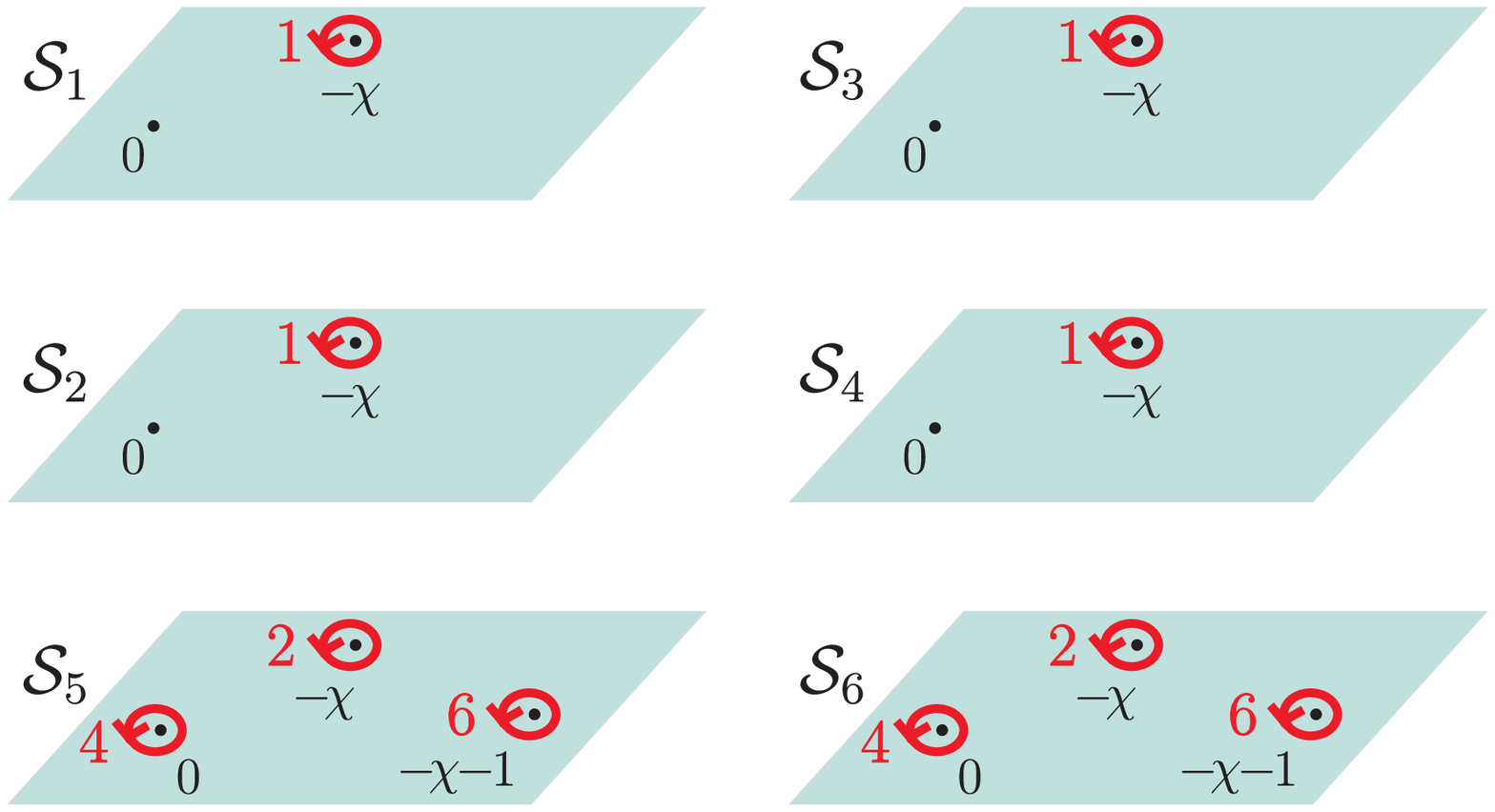}
\\{(b)}
\end{center}
\end{minipage}
\caption{Schematic representation of contours for the coefficients
of the two basis double boxes: (a) the scalar double box,
$\DBox[1]$ (b)
the double box with an irreducible numerator insertion,
$\DBox[\ell_1\cdot k_4]$.
The contours encircle the global poles distributed across the six
kinematical
solutions; the integers next to the contours indicate the winding number.
Both representations are for the choice $u=\frac12$ and $v=1$
in \eqns{eq:heptacut_contours_diag_fam_1}{eq:heptacut_contours_diag_fam_2}.
}\label{ContourFigure}
\end{figure}

The following values,
\begin{eqnarray}
\begin{array}{lll}
a_{1,1} \hspace{1mm}=\hspace{1mm} -2u + v\,,  \hspace{15mm} & 
   a_{2,1} \hspace{1mm}=\hspace{1mm} u\,, & \\
a_{1,2} \hspace{1mm}=\hspace{1mm} -2u + v\,,  \hspace{15mm} & 
   a_{2,2} \hspace{1mm}=\hspace{1mm} u\,, & \\
a_{1,3} \hspace{1mm}=\hspace{1mm} -2u + v\,,  \hspace{15mm} & 
   a_{2,3}\hspace{1mm}=\hspace{1mm} u\,, & \\
a_{1,4} \hspace{1mm}=\hspace{1mm} -2u + v\,,  \hspace{15mm} & 
   a_{2,4} \hspace{1mm}=\hspace{1mm} u\,, & \\
a_{1,5} \hspace{1mm}=\hspace{1mm} 2u - v\,,   \hspace{15mm} & 
   a_{2,5} \hspace{1mm}=\hspace{1mm} v\,,  \hspace{15mm} & 
   a_{3,5} \hspace{1mm}=\hspace{1mm} 2u\,, \\
a_{1,6} \hspace{1mm}=\hspace{1mm} 2u - v\,,   \hspace{15mm} & 
   a_{2,6} \hspace{1mm}=\hspace{1mm} v\,,  \hspace{15mm} & 
   a_{3,6} \hspace{1mm}=\hspace{1mm} 2u\,, 
\end{array} \label{eq:heptacut_contours_diag_fam_1}
\end{eqnarray}
(where $u,v$ are real parameters)
solve all the constraint 
equations~(\ref{Levi-CivitaConstraintFinal}, \ref{eq:heptacut_contour_eq_6}), 
and also set the heptacut of the basis integral
$\DBox[\ell_1\cdot k_4]$ to zero, thereby allowing us
to extract the coefficient of the
first basis integral, $\DBox[1]$.  We will call a specific choice of contours
weighted by these values $P_1$, leaving the dependence on $u$ and $v$
implicit.
A particularly simple solution is
given by $u=\frac{1}{2}$ and $v=1$.  This choice is illustrated schematically
in \fig{ContourFigure}(a).

Similarly, the following values,
\begin{eqnarray}
\begin{array}{lll}
a_{1,1} \hspace{1mm}=\hspace{1mm} -2u + v\,,  \hspace{15mm} & 
   a_{2,1} \hspace{1mm}=\hspace{1mm} u\,, & \\
a_{1,2} \hspace{1mm}=\hspace{1mm} -2u + v\,,  \hspace{15mm} & 
   a_{2,2} \hspace{1mm}=\hspace{1mm} u\,, & \\
a_{1,3} \hspace{1mm}=\hspace{1mm} -2u + v\,,  \hspace{15mm} & 
   a_{2,3} \hspace{1mm}=\hspace{1mm} u\,, & \\
a_{1,4} \hspace{1mm}=\hspace{1mm} -2u + v\,,  \hspace{15mm} & 
   a_{2,4} \hspace{1mm}=\hspace{1mm} u\,, & \\
a_{1,5} \hspace{1mm}=\hspace{1mm} 6u - v\,,   \hspace{15mm} & 
   a_{2,5} \hspace{1mm}=\hspace{1mm} v\,,  \hspace{15mm} & 
   a_{3,5} \hspace{1mm}=\hspace{1mm} 6u\,, \\
a_{1,6} \hspace{1mm}=\hspace{1mm} 6u - v\,,   \hspace{15mm} & 
   a_{2,6} \hspace{1mm}=\hspace{1mm} v\,,  \hspace{15mm} & 
   a_{3,6} \hspace{1mm}=\hspace{1mm} 6u\,, 
\end{array} \label{eq:heptacut_contours_diag_fam_2}
\end{eqnarray}
(where again $u,v$ are real parameters)
solve all the constraint 
equations~(\ref{Levi-CivitaConstraintFinal}, \ref{eq:heptacut_contour_eq_6}), 
 sets to zero the heptacut of the basis
integral $\DBox[1]$, and thereby extracts the coefficient
of $\DBox[\ell_1\cdot k_4]$.  
We will call a specific choice of contours
weighted by these values $P_2$, again leaving the dependence on $u$ and $v$
implicit.
The choice
$u=\frac{1}{2}$ and $v=1$ again gives a particularly simple solution.
It is illustrated schematically
in \fig{ContourFigure}(b).

Before turning to the extraction procedure, we may observe that
the four-dimensional heptacuts do not suffice to extract
information about the coefficients beyond $\Ord(\e^0)$.  The problem
is that we can find non-vanishing linear combinations of tensor integrals
whose heptacut integrand vanishes identically for all six solutions.
As a result, not only do integrals over all 
contours
$T_{j,a}$ vanish, but even integrals constructed by multiplying
the integrand by an arbitrary
function of the remaining degree of freedom $z$ would vanish.  
We call such linear combinations {\it magic\/}.  Examples of magic
combinations include,
\begin{eqnarray}
M_1 &=& \DBox[2,2] + \frac{s_{12}}{2} \DBox[2,1] + \frac{s_{12}}{2} \DBox[1,2]
- \chi \left( \frac{s_{12}}{2} \right)^2 \DBox[1,1] \label{eq:magic_linear_comb}\,,\\
M_2 &=& \DBox[3,2] + \frac{s_{12}}{2} \DBox[3,1] +
\frac{s_{12}}{2} \DBox[2,2] - \chi \left( \frac{s_{12}}{2}
\right)^2 \DBox[2,1] \,, \label{eq:magic_linear_comb_2}
\end{eqnarray}
where the abbreviated notation $\DBox[m,n]$ is defined by,
\begin{equation}
\DBox[m,n] \equiv \DBox[(\ell_1 \cdot k_4)^m (\ell_2 \cdot k_1)^n] \, .
\end{equation}
The magic combinations do not vanish, but both coefficients of
master integrals are of $\Ord(\eps)$ after use of IBP reduction equations.

\section{Integral Coefficients}
\label{ExtractionSection}

With solutions to the constraint equations that also isolate specific
basis integrals in hand, we can write down a procedure for computing
the coefficients of the integrals in the master 
equation~(\ref{TwoLoopBasicEquation}).  To do so, we apply the
augmented heptacuts to the left-hand side of the master equation.  
The basic heptacut will break apart the two-loop amplitude into
a product of six on-shell tree amplitudes, one for each vertex in the
double box.  We will be left with the integral over the $z$ contour.
On the right-hand side, we have the two basis integrals~(\ref{BasisIntegrals})
 chosen earlier.  Here, apply the augmented heptacut, and perform
all integrations.  This gives us the relation,
\begin{equation}
\frac{1}{128} \left( \frac{2}{s_{12}} \right)^3 \sum_{i=1}^6
\oint_{\Gamma_i} \frac{dz}{z (z+\chi)} (-i) \prod_{j=1}^6
A_j^\mathrm{tree}(z) = \frac{c_1}{16 \chi s_{12}^3} \sum_{j=1}^6
(a_{1,j} - a_{2,j}) + \frac{c_2}{32 s_{12}^2} \sum_{j\neq1,3} a_{1,j}\,.
\label{eq:heptacut_of_2-loop_amplitude_ansatz}
\end{equation}
In this equation, the product of amplitudes arises from a factor
of a tree-level amplitude at each vertex of the double box with
all seven propagators cut.

As explained in the previous section,
through a judicious choice of contours, we can make the coefficient
of $c_2$ in this equation vanish, or alternatively the coefficient
of $c_1$ vanish.  This would then allow us to solve for $c_1$ and $c_2$,
respectively.  We gave such choices 
in \eqns{eq:heptacut_contours_diag_fam_1}{eq:heptacut_contours_diag_fam_2}.
Using them, 
we can write an expression for $c_1$,
{\vskip 3mm} \noindent
\framebox[160mm]{\parbox[c][16mm]{150mm}{%
\begin{equation} 
c_1 \hspace{1mm}=\hspace{1mm} 
  \frac{i\chi}{8 u}  \oint_{P_1} \frac{dz}{z (z+\chi)} 
      \prod_{j=1}^6 A_j^\mathrm{tree}(z)
\label{eq:extraction_formula_for_c_1}\,,
\end{equation}}}
{\vskip 3mm}
\noindent and for $c_2$,
{\vskip 2mm} \noindent
\framebox[160mm]{\parbox[c][16mm]{150mm}{%
\begin{equation} 
c_2 \hspace{1mm}=\hspace{1mm} 
  - \frac{i}{4 s_{12} u}  \oint_{P_2} \frac{dz}{z(z+\chi)} 
            \prod_{j=1}^6 A_j^\mathrm{tree}(z)
\label{eq:extraction_formula_for_c_2} \: .
\end{equation}}}
{\vskip 3mm} \noindent 
The right-hand sides of these equations must be summed over
possible helicity and particle-species assignments.
The explicit
integration is understood
to be over the $z$ component of $P_1$ and $P_2$ respectively, 
with the integrations over the other $\alpha_i$ and $\beta_i$
implicit in the solutions $\Sol_j$, and 
with the
dependence of $P_j$ on the parameters $u$ and $v$ left implicit.
The formul\ae{}
(\ref{eq:extraction_formula_for_c_1})
and~(\ref{eq:extraction_formula_for_c_2})
represent the central result of this paper. They are valid for any
gauge theory, and indeed for any amplitude satisfying the power-counting
rules of gauge theory.
With the notation,
\begin{equation}
V\big|_{\Sol_1+\Sol_2-\Sol_3} \equiv
V\big|_{\Sol_1}
+V\big|_{\Sol_2}
-V\big|_{\Sol_3}\,,
\end{equation}
we can write out these formul\ae{} more explicitly,
\begin{eqnarray}
c_1 &=& \frac{(v-2u)i}{8 u} 
 \Res_{z=0} \frac1{z} \left.\prod_{j=1}^6 A_j^\mathrm{tree}(z)
   \right|_{\Sol_1+\Sol_2+\Sol_3+\Sol_4-\Sol_5-\Sol_6}
  -\frac{i v}{8 u} \Res_{z=-\chi} 
    \frac1{z+\chi} \left.\prod_{j=1}^6 A_j^\mathrm{tree}(z)
   \right|_{\Sol_5+\Sol_6}
\nn\\ &&\hskip 5mm
  -\frac{i}{8} \Res_{z=-\chi} 
    \frac1{z+\chi} \left.\prod_{j=1}^6 A_j^\mathrm{tree}(z)
   \right|_{\Sol_1+\Sol_2+\Sol_3+\Sol_4}
  +\frac{i\chi}{4(1+\chi)} \Res_{z=-\chi-1} 
    \left.\prod_{j=1}^6 A_j^\mathrm{tree}(z)
   \right|_{\Sol_5+\Sol_6}\,,
\nn\\
c_2 &=& -\frac{(v-2u)i}{4 s_{12} u\chi} 
 \Res_{z=0} \frac1{z} \left.\prod_{j=1}^6 A_j^\mathrm{tree}(z)
   \right|_{\Sol_1+\Sol_2+\Sol_3+\Sol_4}
  -\frac{(6u-v)i}{4 s_{12} u\chi} 
 \Res_{z=0} \frac1{z} \left.\prod_{j=1}^6 A_j^\mathrm{tree}(z)
   \right|_{\Sol_5+\Sol_6}
\nn\\ &&\hskip 5mm
  +\frac{i}{4 s_{12} \chi} \Res_{z=-\chi} 
    \frac1{z+\chi} \left.\prod_{j=1}^6 A_j^\mathrm{tree}(z)
   \right|_{\Sol_1+\Sol_2+\Sol_3+\Sol_4}
  +\frac{i v}{4 s_{12} u\chi} \Res_{z=-\chi} 
    \frac1{z+\chi} \left.\prod_{j=1}^6 A_j^\mathrm{tree}(z)
   \right|_{\Sol_5+\Sol_6}
\nn\\ &&\hskip 5mm
  -\frac{3 i}{2 s_{12} (1+\chi)} \Res_{z=-\chi-1} 
    \left.\prod_{j=1}^6 A_j^\mathrm{tree}(z)
   \right|_{\Sol_5+\Sol_6}\,.
\label{ExplicitExtractionFormulae}
\end{eqnarray}

These formul\ae{} are not manifestly independent of the choice of contour,
but the constraint equations ensure that they are.  We will see
explicit examples in the next section.  Of course, the independence
of the final result of the choice of contour does not mean 
that the results at intermediate
steps are independent; certain choices of contour may in fact simplify
analytic or numerical calculations.  We have already seen hints of this
in the choices of $P_1$ and $P_2$, where some values of $u$ and $v$
will require evaluation of fewer contours, and hence possibly fewer
numerical evaluations if the formul\ae{} 
(\ref{eq:extraction_formula_for_c_1})
and~(\ref{eq:extraction_formula_for_c_2}) are used in a numerical
setting.

\begin{figure}[!h]
\begin{center}
\includegraphics[angle=0, width=0.4\textwidth]{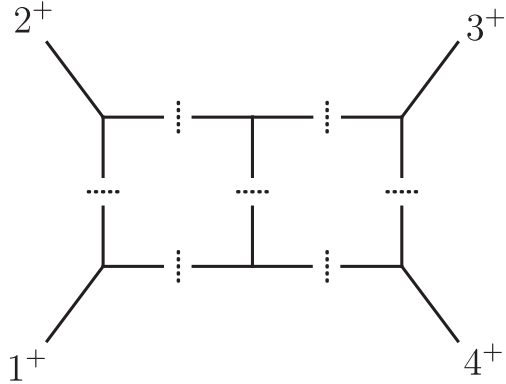}
\caption{Graphical representation of $\left. \prod_{j=1}^6
A_j^\mathrm{tree} (z)\right|_{D=4}$ for the all-plus QCD amplitude.
}\label{AllPlus}
\end{center}
\end{figure}

At one loop, one can choose a basis so that integral coefficients 
are independent of the dimensional regulator $\eps$, and four-dimensional
cuts suffice to compute all of them.  (Computing the rational terms
requires use of $D$-dimensional cuts.)  At two loops, the coefficients
of integral reductions, and hence generally of integrals in
\eqn{TwoLoopBasicEquation}, will depend explicitly on $\eps$. 
In particular, $c_1$ and $c_2$ above will depend explicitly on
$\eps$. In
general, this dependence cannot be extracted from four-dimensional
heptacuts alone, because of the vanishing of
magic combinations discussed in 
\Sect{ConstraintSection}.  We can also see the need for cuts
beyond four dimensions, or considerably relaxing some of the heptacut
conditions, by considering the two-loop all-plus amplitude,
$A_4^\twoloop(+{}+\!{}+{}+)$, computed in ref.~\cite{Bern:2000dn}.
In this case, the product of tree amplitudes in 
\eqns{eq:extraction_formula_for_c_1}{eq:extraction_formula_for_c_2}
will necessarily vanish in four dimensions, because there is no 
assignment of internal helicities in \fig{AllPlus}
that will leave all three-point
amplitudes non-vanishing.  The same observation still holds if we relax
some of the cut conditions, examining hexacuts or pentacuts.

\section{Examples}
\label{ExamplesSection}

In this section, we apply the formalism developed in previous sections
to several examples of two-loop four-point amplitudes.  We use
the master formul\ae{}~(\ref{eq:extraction_formula_for_c_1})
and~(\ref{eq:extraction_formula_for_c_2}) to compute the
coefficients to $\Ord(\eps^0)$ of the two double box basis integrals,
$\DBox[1]$ and $\DBox[\ell_{1}\cdot k_{4}]$.  We consider 
three different contributions to four-gluon amplitudes
in supersymmetric theories with ${\cal N}=4,2,1$ supersymmetries:
the $s$- and $t$-channel contributions to $A_4^\twoloop(1^-,2^-,3^+,4^+)$,
and the $s$-channel contributions to $A_4^\twoloop(1^-,2^+,3^-,4^+)$.
(The $t$-channel contributions to the latter amplitudes
can be obtained by relabeling the arguments of the $s$-channel contribution.)

We will express the results as multiples of the tree-level four-point
amplitudes,
\begin{equation}
A^\mathrm{tree}_{--++} = \frac{i\langle 12\rangle^3}{\langle
23\rangle \langle 34 \rangle \langle 41 \rangle} \,,
\label{eq:definition_of_Atree(1-2-3+4+)}
\end{equation}
and
\begin{equation}
A^\mathrm{tree}_{-+-+} = \frac{i\langle 13\rangle^4}{\langle
12\rangle \langle 23\rangle \langle 34 \rangle \langle 41 \rangle}
\, . \label{eq:definition_of_Atree(1-2+3-4+)}
\end{equation}

\begin{figure}[!h]
\begin{center}
\includegraphics[angle=0, width=0.4\textwidth]{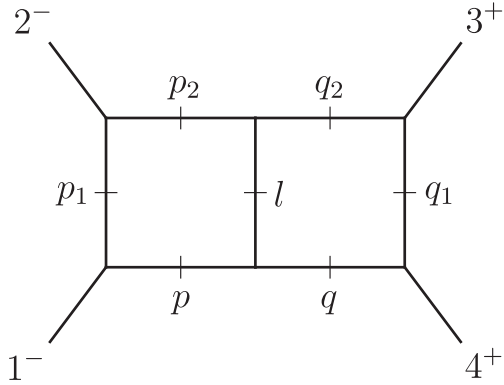}
\end{center}
\caption{The labeling of internal momenta
used in \Sect{ExamplesSection}, here shown for
the $s$-channel contribution to $A_4^\twoloop(1^-,2^-,3^+,4^+)$.}
\label{ExamplesLabeling}
\end{figure}

In this section, it will be convenient to have a label for each
cut propagator in the double box.  Accordingly, we adopt a different
labeling from previous sections.  It is displayed in \fig{ExamplesLabeling}.

\subsection{The $s$-channel contribution to $A_4^\twoloop(1^-,2^-,3^+,4^+)$}
\vskip 10pt

For this contribution, shown above in \fig{ExamplesLabeling},
the helicities of the external states allow only gluons to propagate
in either loop.  For this reason, we will get the same result
independent of the number of supersymmetries.
We find that for all six solutions to the on-shell equations,
\begin{equation}
\prod_{j=1}^6 A_j^\mathrm{tree} \hspace{1mm}=\hspace{1mm} -i
s_{12}^2 s_{23} A^\mathrm{tree}_{--++}\,.
\label{eq:1-2-3+4+_heptacut}
\end{equation}
We can then use \eqn{eq:extraction_formula_for_c_1} (or
equivalently the first equation in \eqn{ExplicitExtractionFormulae})
to obtain,
\begin{eqnarray}
c_1 &=& -i s_{12}^2 s_{23} A^\mathrm{tree}_{--++}\left(
\frac{(v-2u)i}{4 u}   -\frac{i v}{4 u}   -\frac{i}{2} \right)\nn\\
&=& -s_{12}^2 s_{23} A^\mathrm{tree}_{--++}\,;
\label{eq:result_for_1-2-3+4+_1}
\end{eqnarray}
and \eqn{eq:extraction_formula_for_c_2} (or equivalently
the second equation in \eqn{ExplicitExtractionFormulae})
to obtain,
\begin{eqnarray}
c_2 &=& -i s_{12}^2 s_{23} A^\mathrm{tree}_{--++}\left(
-\frac{(v-2u)i}{s_{12} u\chi} 
  -\frac{(6u-v)i}{2 s_{12} u\chi} 
  +\frac{i}{s_{12} \chi}
  +\frac{i v}{2 s_{12} u\chi} \right)\nn\\
&=& 0\,.
\label{eq:result_for_1-2-3+4+_2}
\end{eqnarray}
We see that the dependence on the parameters $u$ and $v$ has
disappeared, as expected.
In the ${\cal N}=4$ theory, these turn out to be the exact coefficients;
in theories with fewer supersymmetries, there are additional
terms of $\Ord(\eps)$ in these coefficients.

\subsection{The $t$-channel contribution to $A_4^\twoloop(1^-,2^-,3^+,4^+)$}
\vskip 10pt

\begin{figure}[!h]
\begin{center}
\includegraphics[angle=0, width=0.4\textwidth]{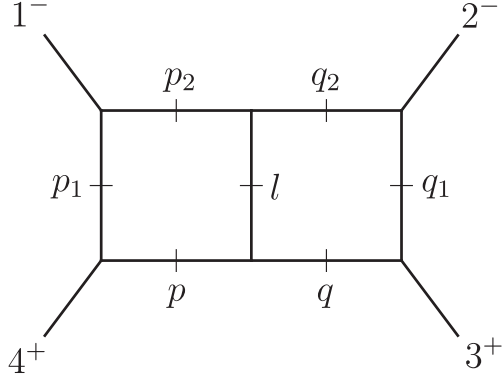}
\end{center}
\caption{The heptacut for the $t$-channel contribution to $A_4^\twoloop(1^-,2^-,3^+,4^+)$.}
\label{tChannelHeptacut}
\end{figure}

We turn next to the computation of the coefficients in the $t$-channel
contribution to the same amplitude considered in the previous section.
The heptacut for this contribution is shown in \fig{tChannelHeptacut}.
In applying the formul\ae{} for the coefficients, we have cyclicly
permuted
the external momentum arguments, $(1,2,3,4)\rightarrow (4,1,2,3)$, so
that we must replace $\chi\rightarrow \chi^{-1}$.  Otherwise, they
are of course unchanged.

\begin{figure}[!ht]
\begin{minipage}[b]{0.45\linewidth}
\begin{center} \includegraphics[scale=0.95]{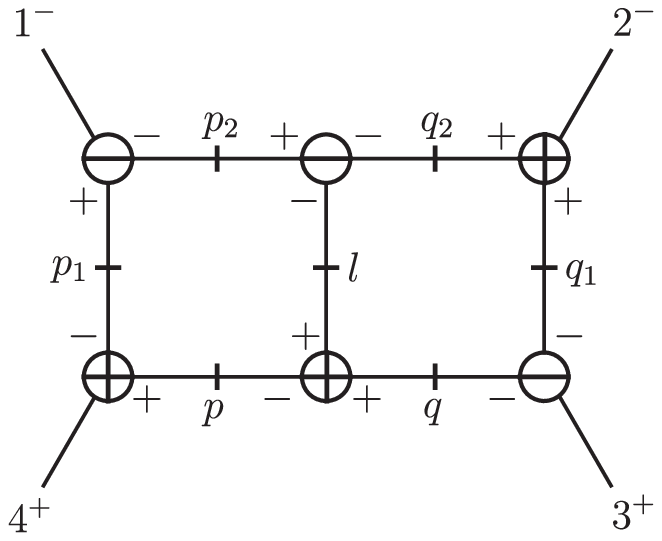}
\\{(a)}
\end{center}
\end{minipage}
\begin{minipage}[b]{0.45\linewidth}
\begin{center} \includegraphics[scale=0.95]{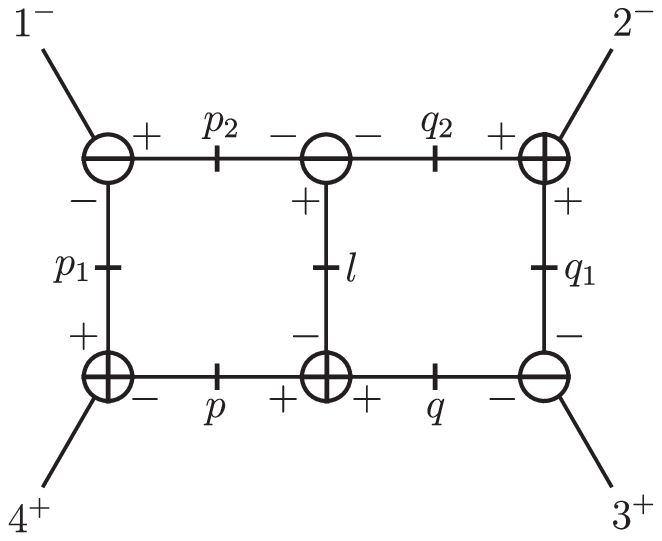}
\\{(b)}
\end{center}
\end{minipage}
\caption{The two distinct assignments of internal helicities 
in solution $\Sol_2$ for
the $t$-channel double-box contributions to $A_4^\twoloop(1^-,2^-,3^+,4^+)$:
(a) configuration A and (b) configuration B.}
\label{TwoConfigurations}
\end{figure}

In this contribution, 
computing the required products of tree amplitudes is 
more involved, and the 
computation also requires sums over supermultiplets
of states propagating in the loops.
  As an example, we work through the computation of
the product in solution $\Sol_2$.  We have two possible
helicity assignments for the internal lines, shown in
\fig{TwoConfigurations}.  For gluon internal lines, we multiply
the amplitudes at the six vertices to obtain,
\begin{equation}
\prod_{j=1}^6 A_j^\mathrm{tree, \hspace{0.7mm} gluon}
\hspace{2mm}=\hspace{2mm} -\frac{1}{\Delta} \times
\left\{
\begin{array}{ll}
A^4 & \mbox{for configuration A}\,, \\
B^4 & \mbox{for configuration B}\,,
\end{array}\right. \label{eq:4+1-2-3+_kin_sol_1gluon_amplitudes}
\end{equation}
where
\begin{eqnarray}
A^4 \hspace{-1mm}&=&\hspace{-1mm} \big( 
\spb{p}.4 \spa1.{p_2} \spa{q_2}.{l}\spb{l}.{q}
 \spa{q}.{q_1}\spb{q_1}.{q_2} \big)^4\,, \nn\\
B^4 \hspace{-1mm}&=&\hspace{-1mm} \big( 
\spb4.{p_1} \spa{p_1}.1 \spa{p_2}.{q_2}\spb{q}.{p}
 \spa{q}.{q_1}\spb{q_1}.{q_2} \big)^4\,,
\label{eq:4+1-2-3+_gluon_index_line_both}\\
\Delta \hspace{-1mm}&=&\hspace{-1mm} 
\spb{p}.4 \spb4.{p_1} \spb{p_1}.{p} \spa{p_1}.1\spa1.{p_2}\spa{p_2}.{p_1}
\spa{p_2}.{q_2}\spa{q_2}.{l}\spa{l}.{p_2}\spb{l}.{q}\spb{q}.{p}\spb{p}.{l}
   \nn \\
&\phantom{=}& \hspace{5mm} \times \spa{q}.{q_1}\spa{q_1}.3\spa3.{q}
\spb{q_1}.{q_2} \spb{q_2}.2 \spb2.{q_1}\,,\nn
\end{eqnarray}
and the minus sign in \eqn{eq:4+1-2-3+_kin_sol_1gluon_amplitudes}
comes from the factor of $i$ in each $A_j^\mathrm{tree,
\hspace{0.7mm} gluon}$.

The helicity assignments of the internal lines allow only gluons to
propagate in the right ($q$) loop, whereas the entire supersymmetric
multiplet of states can propagate in the left ($p$) loop. For
$\mathcal{N}=4$ super Yang-Mills, the sum over states yields,
\begin{equation}
\sum_{{\mathcal{N}=4 \atop \mathrm{multiplet}}} \left.
\prod_{j=1}^6 A_j^\mathrm{tree} \right|_{\Sol_2} \hspace{2mm} =
\hspace{2mm} - \frac{(A+B)^4}{\Delta} \: .
\label{eq:4+1-2-3+_heptacut_sol_1_N=4_result_formal}
\end{equation}
On the other hand, from refs.~\cite{Bern:1997nh}
and~\cite{Octacut} we know that in the ${\cal N}=4$ theory,
\begin{equation}
\sum_{\mathcal{N}=4 \atop \mathrm{multiplet}} \left.
\prod_{j=1}^6 A_j^\mathrm{tree} \right|_{\Sol_2} \hspace{2mm} =
\hspace{2mm} -i s_{12} s_{23}^2 A^\mathrm{tree}_{--++} \,.
\label{eq:4+1-2-3+_heptacut_sol_1_N=4_result_final}
\end{equation}
As a calculational shortcut, we use the equality of
the expressions in \eqns{eq:4+1-2-3+_heptacut_sol_1_N=4_result_formal}
{eq:4+1-2-3+_heptacut_sol_1_N=4_result_final} to fix the relative
sign of $A$ and $B$ in
\eqn{eq:4+1-2-3+_gluon_index_line_both}.
(Of course, the relative signs can also be determined {\it a priori\/}, 
without reference to results in the literature, 
by carefully tracking the direction ---
incoming or outgoing --- of the momenta at a given vertex and
using the analytic continuation rule that changing the sign of
a momentum, $p_i \to -p_i$, is effected by changing the sign of
the holomorphic spinor \cite{Elvang:2008na}: $\lambda_i^\alpha \to
-\lambda_i^\alpha$ while $\widetilde{\lambda}_i^{\dot{\alpha}} \to
\widetilde{\lambda}_i^{\dot{\alpha}}$.)  One finds,
\begin{eqnarray}
A \hspace{-1mm}&=&\hspace{-1mm} 
\spb{p}.4 \spa1.{p_2} \spa{q_2}.{l}\spb{l}.{q}
 \spa{q}.{q_1}\spb{q_1}.{q_2}\,, \nn\\
B \hspace{-1mm}&=&\hspace{-1mm} -
\spb4.{p_1} \spa{p_1}.1 \spa{p_2}.{q_2}\spb{q}.{p}
 \spa{q}.{q_1}\spb{q_1}.{q_2} \, .
\end{eqnarray}
Ref.~\cite{Bern:2009xq} teaches us that the sum over the
$\mathcal{N}=4,2,1,0$ multiplet of states is related to the
$\mathcal{N}=4$ state sum via
\begin{equation}
\sum_{{\mathrm{SUSY} \atop \mathrm{multiplet}}}
\prod_{j=1}^6 A_j^\mathrm{tree} \hspace{1.5mm}=\hspace{1.5mm}
\frac{(A+B)^\mathcal{N} (A^{4-\mathcal{N}} +
B^{4-\mathcal{N}})}{(A+B)^4} \left( 1- \textstyle{\frac{1}{2}}
\delta_{\mathcal{N},4}\right) 
\sum_{{\mathcal{N}=4 \atop \mathrm{multiplet}}}
    \prod_{j=1}^6 A_j^\mathrm{tree}\,,
\label{eq:relation_between_SUSY_state_sums}
\end{equation}
so that the sum over the supersymmetric multiplet of states can be
calculated from the gluonic contributions alone (indeed, recall
that $A$ and $B$ in
\eqn{eq:4+1-2-3+_gluon_index_line_both}
were obtained from the product of purely gluonic amplitudes corresponding to
configurations A and B, respectively).

We can simplify the expression for the ratio between the
supersymmetric state sums in
\eqn{eq:relation_between_SUSY_state_sums} by factoring out as many
common factors of $A$ and $B$ as possible (exploiting momentum
conservation fully). Setting $A = \alpha F$ and $B = \beta F$,
for $\mathcal{N}=4,2,1$  the ratio appearing 
in \eqn{eq:relation_between_SUSY_state_sums}
simplifies to
\begin{eqnarray}
R \hspace{-1mm}&=&\hspace{-1mm} \frac{(\alpha+\beta)^\mathcal{N}
(\alpha^{4-\mathcal{N}} +
\beta^{4-\mathcal{N}})}{(\alpha+\beta)^4} \left( 1-
\textstyle{\frac{1}{2}} \delta_{\mathcal{N},4}\right)
\hspace{1mm}=\hspace{1mm} \frac{\left( \alpha^{4-\mathcal{N}} +
\beta^{4-\mathcal{N}} \right) \left( 1- \textstyle{\frac{1}{2}}
\delta_{\mathcal{N},4}\right)}{(\alpha+\beta)^{4-\mathcal{N}}} \label{eq:4+1-2-3+_simplyfing_R_formally_1}\\
&=& \hspace{-1mm} 1 - (4-\mathcal{N}) \left( \frac{\alpha}{\alpha
+ \beta} \right) + (4-\mathcal{N}) \left( \frac{\alpha}{\alpha +
\beta} \right)^2\,. \label{eq:4+1-2-3+_simplyfing_R_formally_2}
\end{eqnarray}
where the last equality holds only
for $\mathcal{N}=4,2,1$; it can be obtained by
expanding the numerator $\left( \alpha^{4-\mathcal{N}} +
\beta^{4-\mathcal{N}} \right) \left( 1- \textstyle{\frac{1}{2}}
\delta_{\mathcal{N},4}\right)$ in
\eqn{eq:4+1-2-3+_simplyfing_R_formally_1} in $\beta$ around
$-\alpha$.

In the case at hand, we can use momentum conservation
($l = p_2 + q_2$ and $p_1 = p - k_4$) to rewrite $A$ and $B$ 
as follows,
\begin{eqnarray}
A \hspace{-1mm}&=&\hspace{-1mm} 
\spb{p}.4 \spa1.{p_2} \spa{q_2}.{p_2}\spb{p_2}.{q}
 \spa{q}.{q_1}\spb{q_1}.{q_2}\,, \nn\\
B \hspace{-1mm}&=&\hspace{-1mm} -
\spb4.{p} \spa{p}.1 \spa{p_2}.{q_2}\spb{q}.{p}
 \spa{q}.{q_1}\spb{q_1}.{q_2} \,,
\end{eqnarray}
and identify,
\begin{eqnarray}
\alpha &=& \spa1.{p_2} \spb{p_2}.{q}\,,\nn\\
\beta &=& -\spb{q}.{p}\spa{p}.1\,,\\
F &=& \spb{p}.4 \spa{q_2}.{p_2}
      \spa{q}.{q_1}\spb{q_1}.{q_2}\,. \nn
\end{eqnarray}
Momentum conservation implies that 
$\alpha + \beta = -\spa1.4\spb4.{q}$, and thus,
\begin{equation}
\frac{\alpha}{\alpha + \beta} = 
- \frac{\spa1.{p_1}\spb{p_1}.{q}}{\spa1.4\spb4.{q}} 
= - \frac{\spa1.{p_1}\spb{p_1}.3}{\spa1.4\spb4.3}\,,
\end{equation}
where the second equality uses the proportionality of antiholomorphic
spinors, $\tlambda_q \propto \tlambda_3$.  (This proportionality
holds only for some of the other six solutions $\Sol_i$ in addition
to $\Sol_2$.)
The ratio thus simplifies
to,
\begin{equation}
R = 1 + (4-\mathcal{N}) \left( 
  \frac{\spa1.{p_1}\spb{p_1}.3}{\spa1.4\spb4.3}\right) + (4-\mathcal{N}) \left(
  \frac{\spa1.{p_1}\spb{p_1}.3}{\spa1.4\spb4.3}\right)^2\,.
\end{equation}
We can solve for the explicit values of the cut momenta using
the parametrization~(\ref{TwoLoopParametrization}) with the
external momenta cyclicly permuted (for the $t$-channel configuration),
and using the on-shell values defining $\Sol_2$ given in \eqns{AlphaBetaEquations}{Solutions1and2}.
We find,
\begin{equation}
p_1^\mu \equiv p^\mu - k_4^\mu = \frac{s_{14} z}{2\spa4.3\spb3.1} 
\sand4.{\gamma^\mu}.1\,,
\end{equation}
so that,
\begin{eqnarray}
\spa1.{p_1}\spb{p_1}.3 &=& \sand1.{\gamma^\mu}.3 p_{1\mu} = 
  \frac{s_{14} z}{2\spa4.3\spb3.1}\sand1.{\gamma^\mu}.3
\sand4.{\gamma_\mu}.1\nn \\
&=& \frac{\spa4.1}{\spa4.3} s_{14} z\,,
\end{eqnarray}
and thus,
\begin{equation}
\frac{\spa1.{p_1}\spb{p_1}.3}{\spa1.4\spb4.3} =
\chi z \, .
\end{equation}
This gives us our final form for the ratio,
\begin{equation}
R = 1 + (4-\mathcal{N}) \chi z + (4-\mathcal{N}) \chi^2 z^2\,,
\end{equation}
and for the product of tree amplitudes,
\begin{equation}
\sum_{{\mathrm{SUSY} \atop \mathrm{multiplet}}}
\prod_{j=1}^6 A_j^\mathrm{tree}\bigg|_{\Sol_2} =
 \hspace{-1mm} -i s_{12} s_{23}^2 A^\mathrm{tree}_{--++} \Big(
1 + (4-\mathcal{N}) \chi z + (4-\mathcal{N}) \chi^2 z^2 \Big)\,.
\label{eq:4+1-2-3+_cut_sol_1_simplified}
\end{equation}

In this solution to the heptacut equations, the supersymmetric
multiplet runs only in one of the loops.  In other solutions (in
particular, $\Sol_6$), the multiplet can run in both loops.  The
treatment of this case is similar but more elaborate.  It turns
out~\cite{Bern:2009xq} that the sum over the multiplet can again be
evaluated purely from the gluonic contributions.  The main difference
is that in this case there are three gluonic contributions $A^4, B^4,
C^4$ (compared to the two in \eqn{eq:4+1-2-3+_gluon_index_line_both}).
One can again fix the relative sign of $B$ and $C$ by insisting that
the $\mathcal{N}=4$ supersymmetric result $- \frac{(A+B+C)^4}{\Delta}$
be equal to \eqn{eq:4+1-2-3+_heptacut_sol_1_N=4_result_final}, and
from the obvious analog of \eqn{eq:relation_between_SUSY_state_sums}
one then finds the results for the supermultiplet sums for
$\mathcal{N}=4,2,1,0$. These expressions can again be simplified as
above.

Summing over all six solutions, 
and plugging the result into our
master formul\ae{}~(\ref{eq:extraction_formula_for_c_1})
and~(\ref{eq:extraction_formula_for_c_2}), taken
with $u=\frac12$ and $v=1$,
we find 
\begin{eqnarray}
c_1 &=& -s_{12} s_{23}^2 A^\mathrm{tree}_{--++} \left(1 +
\frac{1}{4} (1-\delta_{\mathcal{N},4})(4-\mathcal{N})!\, \chi (\chi
+ 1)^{\delta_{\mathcal{N},1}} \right)\,, \nn\\
c_2 &=& \frac{3}{2} s_{23}^2 A^\mathrm{tree}_{--++}
(1-\delta_{\mathcal{N},4})(4-\mathcal{N})!\, \chi (\chi +
1)^{\delta_{\mathcal{N},1}}\,, \label{eq:result_for_4+1-2-3+}
\end{eqnarray}
valid for $\mathcal{N}=4,2,1$. 

\subsection{The $s$-channel contribution to $A_4^\twoloop(1^-,2^+,3^-,4^+)$}
\vskip 10pt

\begin{figure}[!h]
\begin{center}
\includegraphics[angle=0, width=0.4\textwidth]{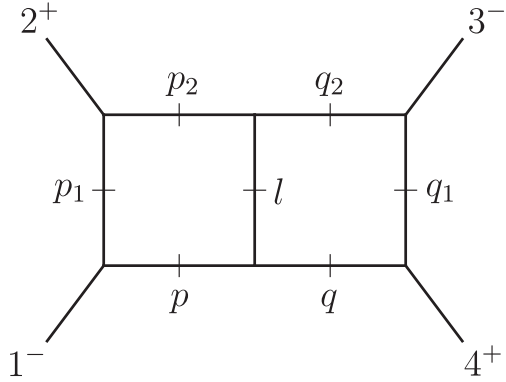}
\end{center}
\caption{The heptacut for the $s$-channel contribution to 
$A_4^\twoloop(1^-,2^+,3^-,4^+)$.}
\label{sChannel-+-+Heptacut}
\end{figure}
The heptacut for the $s$-channel contribution to 
$A_4^\twoloop(1^-,2^+,3^-,4^+)$ is shown in 
\fig{sChannel-+-+Heptacut}.  We will evaluate this contribution
in two different ways, illustrating both the result's independence
of the precise choice of contour, and also illustrating the
potential advantages of a judicious choice of contour in a given calculation.

Rather than using our master formul\ae{}~(\ref{eq:extraction_formula_for_c_1})
and~(\ref{eq:extraction_formula_for_c_2}), 
let us evaluate the augmented heptacut integral for a general contour,
before imposing the constraint equations.  Adding up the contributions
from all six solutions, we find
\begin{eqnarray}
&& \hspace{-11mm} \sum_{i=1}^6 \oint_{\Gamma_i}
\frac{dz}{z(z+\chi)} \prod_{j=1}^6 A_j^\mathrm{tree}(z) = \nn \\
&& -i s_{12}^2 s_{23} A^\mathrm{tree}_{-+-+} \left[ \sum_{j=1}^6
\frac{a_{1,j} -
a_{2,j}}{\chi} - (4 - \mathcal{N}) \frac{a_{1,6} - a_{3,6} - a_{2,5}}{(\chi + 1)^2} \right. \nonumber \\
&\phantom{=}& \left. \hspace{30mm} + \left( \left( 1 -
\textstyle{\frac{1}{2}} \delta_{\mathcal{N},4} \right)
\frac{\chi^{4 - \mathcal{N}} + 1}{(\chi+1)^{4 - \mathcal{N}}} - 1
\right) \left( \frac{a_{1,3} - a_{2,3}}{\chi}
             + \frac{a_{1,4} - a_{2,4}}{\chi} 
             \right) \right] \, . \phantom{aaa}
\label{eq:1-2+3-4+_heptacut_intermediate_step}
\end{eqnarray}

In this expression, we need to impose the constraint equations
in order to restrict the evaluation to a valid contour; and then
we would seek to project onto each basis integral in turn.  Now,
suppose we can find a pair of solutions to the constraint equations which
projects onto the first or second basis integral, respectively, 
{\it and\/} in addition, satisfies 
$a_{1,3} - a_{2,3} + a_{1,4} - a_{2,4}  = 0$.  Using such a contour
would set
the second line of \eqn{eq:1-2+3-4+_heptacut_intermediate_step}
equal to zero and therefore produce a particularly simple
algebraic expression for $c_1$ and $c_2$ directly, without need
for additional simplification.  Choosing $u=\frac13$ and $v=1$
in $P_1$ and $P_2$ gives such a contour.

This gives us the results,
\begin{eqnarray}
c_1 &=& -s_{12}^2 s_{23} A^\mathrm{tree}_{-+-+} \left( 1 -
\frac{3}{4}
(4 - \mathcal{N}) \frac{\chi}{(\chi + 1)^2} \right)\,,\nn \\
c_2 &=& -\frac{3}{2} s_{12} s_{23} A^\mathrm{tree}_{-+-+} \frac{4
- \mathcal{N}}{(\chi + 1)^2}\,,
\label{eq:result_for_c_2_for_1-2+3-4+}
\end{eqnarray}
valid for $\mathcal{N}=4,2,1$.  The $t$-channel contribution can
be obtained by exchanging $s_{12}\leftrightarrow s_{23}$ and
$\chi\rightarrow \chi^{-1}$.

If we compare the
expressions obtained above for the coefficients $c_i$ to
those obtained using the choice
suggested in \Sect{ExtractionSection}, $u=\frac12$ and $v=1$,
we find that the expressions are equal by virtue of the identity
\begin{equation}
\frac{1}{\chi} \left( \left( 1 - \textstyle{\frac{1}{2}}
\delta_{\mathcal{N},4} \right) \frac{\chi^{4 - \mathcal{N}} +
1}{(\chi+1)^{4 - \mathcal{N}}} - 1 \right) =
-\frac{4-\mathcal{N}}{(\chi + 1)^2}\,,
\label{eq:simplifying_identity_for_heptacut}
\end{equation}
valid for $\mathcal{N}=4,2,1$. This identity can of course easily
be proven without reference to the current discussion, but the
point we wish to emphasize is that the flexibility in choosing
contours suggests certain algebraic simplifications which are not
immediately obvious.

The double box coefficients given in eqs.~(\ref{eq:result_for_1-2-3+4+_1}, %
\ref{eq:result_for_1-2-3+4+_2}, \ref{eq:result_for_4+1-2-3+}, %
\ref{eq:result_for_c_2_for_1-2+3-4+})
agree with the $\Ord(\eps^0)$ terms of the corresponding
coefficients, supplied to us by Lance Dixon~\cite{DixonPrivate},
  in the amplitudes computed by 
Bern, De~Freitas, and Dixon~\cite{Bern:2002tk}.

\section{Conclusions}
\label{ConclusionSection}

In this paper, we have taken the first step to extending
the maximal generalized unitarity method to two loops.  Cutting
propagators can be viewed as deforming the original real loop-momentum
contours of integration to contours encircling the global poles of
the integrand.  At two loops, there is a variety of such poles.
We can evaluate the integral along many different linear combinations
of these contours.  However, our choices are restricted by the
requirement that the evaluation along any contour respect the
vanishing of certain insertions of Levi-Civita symbols, as well as of
total derivatives arising from integration-by-parts identities.  We
derived the corresponding constraint equations for the massless double
box, and showed how to use their solutions to obtain simple formul\ae{},
\eqns{eq:extraction_formula_for_c_1}{eq:extraction_formula_for_c_2},
for the coefficients of the two double box basis integrals to leading
order in the dimensional regulator $\eps$.  To derive these
equations, we adopted a parametrization of the loop momenta and
solved explicitly for the maximal cuts, a heptacut in our case,
and identified the additional poles present in the remaining degree
of freedom.

We expect that the approach given in this paper --- parametrize
the basis integrals; solve the on-shell equations; identify the poles
in the remaining degrees of freedom; impose all constraint equations ---
will apply to the full set of integrals required for two-loop amplitudes,
both to the four-dimensional cuts considered here, and more
generally to the $D$-dimensional cuts required for a complete calculation
of the amplitude.

\section*{Acknowledgments}

\vskip -.3 cm 
We are grateful to Henrik Johansson for many useful
discussions and comments on the manuscript.  We also wish to thank
Lance Dixon for sharing unpublished results on the integral
coefficients of the amplitudes computed in ref.~\cite{Bern:2002tk}.
We have also benefited from discussions with Simon Badger, Till
Bargheer, Emil Bjerrum-Bohr, Ruth Britto, Darren Forde, Lisa Freyhult,
Janusz Gluza, Gregory Korchemsky, Donal O'Connell, Mikael Passare and
David Skinner.  KJL thanks the Niels Bohr Institute for hospitality
where part of this work was carried out.  This work is supported by
the European Research Council under Advanced Investigator Grant
ERC--AdG--228301.  KJL gratefully acknowledges financial support from
Kungliga Vetenskaps\-akademien under Project no.{} FOA10V-133.  This
research also used resources of Academic Technology Services at UCLA.




\begin{thebibliography}{99}

\bibitem{LesHouches}
  Z.~Bern {\it et al.}  [NLO Multileg Working Group],
  arXiv:0803.0494 [hep-ph].

\bibitem{UnitarityMethod}
Z.~Bern, L.~J.~Dixon, D.~C.~Dunbar and D.~A.~Kosower,
Nucl.\ Phys.\ B {\bf 425}, 217 (1994)
[hep-ph/9403226];
%
Nucl.\ Phys.\ B {\bf 435}, 59 (1995)
[hep-ph/9409265];\\
%
Z.\ Bern, L.\ J.\ Dixon and D.\ A.\ Kosower,
Ann.\ Rev.\ Nucl.\ Part.\ Sci.\  {\bf 46}, 109 (1996)
[hep-ph/9602280].

\bibitem{Bern:1995db}
Z.~Bern and A.~G.~Morgan,
Nucl.\ Phys.\  B {\bf 467}, 479 (1996)
[arXiv:hep-ph/9511336].

\bibitem{Zqqgg}
Z.~Bern, L.~J.~Dixon and D.~A.~Kosower,
Nucl.\ Phys.\  B {\bf 513}, 3 (1998)
[hep-ph/9708239].

\bibitem{DdimensionalI}
Z.~Bern, L.~J.~Dixon, D.~C.~Dunbar and D.~A.~Kosower,
Phys.\ Lett.\  B {\bf 394}, 105 (1997)
[arXiv:hep-th/9611127].

\bibitem{BCFUnitarity}
R.~Britto, F.~Cachazo and B.~Feng,
Nucl.\ Phys.\  B {\bf 725}, 275 (2005)
[hep-th/0412103].

\bibitem{OtherUnitarity}
R.~Britto, F.~Cachazo and B.~Feng,
Phys.\ Rev.\  D {\bf 71}, 025012 (2005)
[hep-th/0410179];\\
%
S.~J.~Bidder, N.~E.~J.~Bjerrum-Bohr, L.~J.~Dixon and D.~C.~Dunbar,
Phys.\ Lett.\  B {\bf 606}, 189 (2005)
[hep-th/0410296];\\
%
S.~J.~Bidder, N.~E.~J.~Bjerrum-Bohr, D.~C.~Dunbar and W.~B.~Perkins,
Phys.\ Lett.\  B {\bf 612}, 75 (2005)
[hep-th/0502028];\\
%
S.~J.~Bidder, D.~C.~Dunbar and W.~B.~Perkins,
JHEP {\bf 0508}, 055 (2005)
[hep-th/0505249];\\
%
Z.~Bern, N.~E.~J.~Bjerrum-Bohr, D.~C.~Dunbar and H.~Ita,
JHEP {\bf 0511}, 027 (2005)
[hep-ph/0507019];\\
%
N.~E.~J.~Bjerrum-Bohr, D.~C.~Dunbar and W.~B.~Perkins,
0709.2086 [hep-ph].

\bibitem{Bootstrap}
Z.~Bern, L.~J.~Dixon and D.~A.~Kosower,
Phys.\ Rev.\  D {\bf 73}, 065013 (2006)
[hep-ph/0507005].

\bibitem{BCFCutConstructible}
R.~Britto, E.~Buchbinder, F.~Cachazo and B.~Feng,
Phys.\ Rev.\  D {\bf 72}, 065012 (2005)
hep-ph/0503132];\\
%
R.~Britto, B.~Feng and P.~Mastrolia,
Phys.\ Rev.\  D {\bf 73}, 105004 (2006)
[hep-ph/0602178];\\
%
P.~Mastrolia,
Phys.\ Lett.\  B {\bf 644}, 272 (2007)
[hep-th/0611091].

\bibitem{BMST}
A.~Brandhuber, S.~McNamara, B.~J.~Spence and G.~Travaglini,
JHEP {\bf 0510}, 011 (2005)
[hep-th/0506068].

\bibitem{OPP}
G.~Ossola, C.~G.~Papadopoulos and R.~Pittau,
Nucl.\ Phys.\  B {\bf 763}, 147 (2007)
[hep-ph/0609007].

\bibitem{OnShellReview}
Z.~Bern, L.~J.~Dixon and D.~A.~Kosower,
Annals Phys.\  {\bf 322}, 1587 (2007)
[0704.2798 [hep-ph]].

\bibitem{Forde}
D.~Forde,
Phys.\ Rev.\  D {\bf 75}, 125019 (2007)
[0704.1835 [hep-ph]].
%
\bibitem{Badger}
S.~D.~Badger,
JHEP {\bf 0901}, 049 (2009)
[0806.4600 [hep-ph]].

\bibitem{DdimensionalII}
C.~Anastasiou, R.~Britto, B.~Feng, Z.~Kunszt and P.~Mastrolia,
Phys.\ Lett.\  B {\bf 645}, 213 (2007)
[hep-ph/0609191];
%
JHEP {\bf 0703}, 111 (2007)
[hep-ph/0612277];\\
W.~T.~Giele, Z.~Kunszt and K.~Melnikov,
JHEP {\bf 0804}, 049 (2008)
[arXiv:0801.2237 [hep-ph]].

\bibitem{BFMassive}
R.~Britto and B.~Feng,
Phys.\ Rev.\  D {\bf 75}, 105006 (2007)
[hep-ph/0612089].
JHEP {\bf 0802}, 095 (2008)
[0711.4284 [hep-ph]];\\
%
R.~Britto, B.~Feng and P.~Mastrolia,
Phys.\ Rev.\  D {\bf 78}, 025031 (2008)
[arXiv:0803.1989 [hep-ph]];\\
%
R.~Britto, B.~Feng and G.~Yang,
JHEP {\bf 0809}, 089 (2008)
[arXiv:0803.3147 [hep-ph]].

\bibitem{BergerFordeReview}
  C.~F.~Berger and D.~Forde,
  arXiv:0912.3534 [hep-ph].

\bibitem{Bern:2010qa}
  Z.~Bern, J.~J.~Carrasco, T.~Dennen, Y.~T.~Huang and H.~Ita,
  Phys.\ Rev.\  D {\bf 83}, 085022 (2011)
  [arXiv:1010.0494 [hep-th]].

\bibitem{EGK}
R.~K.~Ellis, W.~T.~Giele and Z.~Kunszt,
JHEP {\bf 0803}, 003 (2008)
[0708.2398 [hep-ph]].

\bibitem{BlackHatI}
C.~F.~Berger, Z.~Bern, L.~J.~Dixon, F.~Febres Cordero, D.~Forde, H.~Ita,
D.~A.~Kosower and D.~Ma\^{\i}tre,
Phys.\ Rev.\ D {\bf 78}, 036003 (2008)
[0803.4180 [hep-ph]].

\bibitem{CutTools}
G.~Ossola, C.~G.~Papadopoulos and R.~Pittau,
JHEP {\bf 0803}, 042 (2008)
[arXiv:0711.3596 [hep-ph]].

\bibitem{MOPP}
P.~Mastrolia, G.~Ossola, C.~G.~Papadopoulos and R.~Pittau,
JHEP {\bf 0806}, 030 (2008)
[arXiv:0803.3964 [hep-ph]].

\bibitem{Rocket}
W.~T.~Giele and G.~Zanderighi,
JHEP {\bf 0806}, 038 (2008)
[arXiv:0805.2152 [hep-ph]];\\
R.~K.~Ellis, W.~T.~Giele, Z.~Kunszt, K.~Melnikov and G.~Zanderighi,
JHEP {\bf 0901}, 012 (2009)
[arXiv:0810.2762 [hep-ph]].

\bibitem{BlackHatII}
C.~F.~Berger, Z.~Bern, L.~J.~Dixon, F.~Febres Cordero, D.~Forde, T.~Gleisberg,
H.~Ita, D.~A.~Kosower and D.~Ma\^{\i}tre,
Phys.\ Rev.\ Lett.\  {\bf 102}, 222001 (2009)
[0902.2760 [hep-ph]].

\bibitem{CutToolsHelac}
G.~Bevilacqua, M.~Czakon, C.~G.~Papadopoulos, R.~Pittau and M.~Worek,
JHEP {\bf 0909}, 109 (2009)
[arXiv:0907.4723 [hep-ph]].

\bibitem{Samurai}
P.~Mastrolia, G.~Ossola, T.~Reiter and F.~Tramontano,
JHEP {\bf 1008}, 080 (2010)
[arXiv:1006.0710 [hep-ph]].

\bibitem{WPlus4}
C.~F.~Berger {\it et al.},
Phys.\ Rev.\ Lett.\  {\bf 106}, 092001 (2011)
[arXiv:1009.2338 [hep-ph]].

\bibitem{NGluon}
S.~Badger, B.~Biedermann and P.~Uwer,
Comput.\ Phys.\ Commun.\  {\bf 182}, 1674 (2011)
[arXiv:1011.2900 [hep-ph]].

\bibitem{MadLoop}
V.~Hirschi, R.~Frederix, S.~Frixione, M.~V.~Garzelli, F.~Maltoni and R.~Pittau,
JHEP {\bf 1105}, 044 (2011)
[arXiv:1103.0621 [hep-ph]].

\bibitem{Ellis:1987xu}
  R.~K.~Ellis, I.~Hinchliffe, M.~Soldate and J.~J.~van der Bij,
  Nucl.\ Phys.\  B {\bf 297}, 221 (1988).

\bibitem{Berger:1983yi}
  E.~L.~Berger, E.~Braaten and R.~D.~Field,
  Nucl.\ Phys.\  B {\bf 239}, 52 (1984).

\bibitem{Aurenche:1985yk}
  P.~Aurenche, A.~Douiri, R.~Baier, M.~Fontannaz and D.~Schiff,
  Z.\ Phys.\  C {\bf 29}, 459 (1985).

\bibitem{Bern:2002jx}
  Z.~Bern, L.~J.~Dixon and C.~Schmidt,
  Phys.\ Rev.\  D {\bf 66}, 074018 (2002)
  [arXiv:hep-ph/0206194].

\bibitem{NNLOThreeJet}
 A.~Gehrmann-De Ridder, T.~Gehrmann, E.~W.~N.~Glover and G.~Heinrich,
  JHEP {\bf 0711}, 058 (2007)
  [arXiv:0710.0346 [hep-ph]];\\
  S.~Weinzierl,
  Phys.\ Rev.\ Lett.\  {\bf 101}, 162001 (2008)
  [arXiv:0807.3241 [hep-ph]].

\bibitem{ThreeJetAlphaS}
  G.~Dissertori, A.~Gehrmann-De Ridder, T.~Gehrmann, E.~W.~N.~Glover, G.~Heinrich, G.~Luisoni and H.~Stenzel,
  JHEP {\bf 0908}, 036 (2009)
  [arXiv:0906.3436 [hep-ph]];\\
  G.~Dissertori, A.~Gehrmann-De Ridder, T.~Gehrmann, E.~W.~N.~Glover, G.~Heinrich and H.~Stenzel,
  Phys.\ Rev.\ Lett.\  {\bf 104}, 072002 (2010)
  [arXiv:0910.4283 [hep-ph]].

\bibitem{Bern:1997nh}
Z.~Bern, J.~S.~Rozowsky and B.~Yan,
Phys.\ Lett.\  B {\bf 401}, 273 (1997)
[arXiv:hep-ph/9702424].

\bibitem{ABDK}
  C.~Anastasiou, Z.~Bern, L.~J.~Dixon and D.~A.~Kosower,
  Phys.\ Rev.\ Lett.\  {\bf 91}, 251602 (2003)
  [arXiv:hep-th/0309040].

\bibitem{Bern:2005iz}
  Z.~Bern, L.~J.~Dixon and V.~A.~Smirnov,
  Phys.\ Rev.\  D {\bf 72}, 085001 (2005)
  [arXiv:hep-th/0505205].

\bibitem{Bern:2006vw}
  Z.~Bern, M.~Czakon, D.~A.~Kosower, R.~Roiban and V.~A.~Smirnov,
  Phys.\ Rev.\ Lett.\  {\bf 97}, 181601 (2006)
  [arXiv:hep-th/0604074].

\bibitem{Bern:2006ew}
  Z.~Bern, M.~Czakon, L.~J.~Dixon, D.~A.~Kosower and V.~A.~Smirnov,
  Phys.\ Rev.\  D {\bf 75}, 085010 (2007)
  [arXiv:hep-th/0610248].

\bibitem{Bern:2007ct}
  Z.~Bern, J.~J.~M.~Carrasco, H.~Johansson and D.~A.~Kosower,
  Phys.\ Rev.\  D {\bf 76}, 125020 (2007)
  [arXiv:0705.1864 [hep-th]].

\bibitem{Bern:2008ap}
Z.~Bern, L.~J.~Dixon, D.~A.~Kosower, R.~Roiban, M.~Spradlin, C.~Vergu and A.~Volovich,
Phys.\ Rev.\  D {\bf 78}, 045007 (2008)
[arXiv:0803.1465 [hep-th]].

\bibitem{Bern:2008pv}
  Z.~Bern, J.~J.~M.~Carrasco, L.~J.~Dixon, H.~Johansson and R.~Roiban,
  Phys.\ Rev.\  D {\bf 78}, 105019 (2008)
  [arXiv:0808.4112 [hep-th]].

\bibitem{LeadingSingularity}
F.~Cachazo,
arXiv:0803.1988 [hep-th];\\
F.~Cachazo, M.~Spradlin and A.~Volovich,
Phys.\ Rev.\  D {\bf 78}, 105022 (2008)
[arXiv:0805.4832 [hep-th]].

\bibitem{ArkaniHamed:2009dn}
  N.~Arkani-Hamed, F.~Cachazo, C.~Cheung and J.~Kaplan,
  JHEP {\bf 1003}, 020 (2010)
  [arXiv:0907.5418 [hep-th]].
 
\bibitem{ArkaniHamed:2010kv}
  N.~Arkani-Hamed, J.~L.~Bourjaily, F.~Cachazo, S.~Caron-Huot and J.~Trnka,
  JHEP {\bf 1101}, 041 (2011)
  [arXiv:1008.2958 [hep-th]].

\bibitem{Bern:2010tq}
  Z.~Bern, J.~J.~M.~Carrasco, L.~J.~Dixon, H.~Johansson and R.~Roiban,
  Phys.\ Rev.\  D {\bf 82}, 125040 (2010)
  [arXiv:1008.3327 [hep-th]].

\bibitem{Kosower:2010yk}
D.~A.~Kosower, R.~Roiban and C.~Vergu,
Phys.\ Rev.\  D {\bf 83}, 065018 (2011)
[arXiv:1009.1376 [hep-th]].

\bibitem{ArkaniHamed:2010gh}
  N.~Arkani-Hamed, J.~L.~Bourjaily, F.~Cachazo and J.~Trnka,
  arXiv:1012.6032 [hep-th].

\bibitem{Carrasco:2011mn}
  J.~J.~M.~Carrasco and H.~Johansson,
  arXiv:1106.4711 [hep-th].

\bibitem{Carrasco:2011hw}
  J.~J.~M.~Carrasco and H.~Johansson,
  arXiv:1103.3298 [hep-th].

\bibitem{Bern:2011rj}
  Z.~Bern, C.~Boucher-Veronneau and H.~Johansson,
  arXiv:1107.1935 [hep-th].

\bibitem{Bern:2000dn}
  Z.~Bern, L.~J.~Dixon and D.~A.~Kosower,
  JHEP {\bf 0001}, 027 (2000)
  [arXiv:hep-ph/0001001].

\bibitem{Bern:2002tk}
  Z.~Bern, A.~De Freitas and L.~J.~Dixon,
  JHEP {\bf 0203}, 018 (2002)
  [arXiv:hep-ph/0201161].

\bibitem{BernDeFreitasDixonTwoPhoton}
  Z.~Bern, A.~De Freitas and L.~J.~Dixon,
  JHEP {\bf 0109}, 037 (2001)
  [arXiv:hep-ph/0109078].

\bibitem{Bern:2002zk}
  Z.~Bern, A.~De Freitas, L.~J.~Dixon and H.~L.~Wong,
  Phys.\ Rev.\  D {\bf 66}, 085002 (2002)
  [arXiv:hep-ph/0202271].

\bibitem{Bern:2003ck}
  Z.~Bern, A.~De Freitas and L.~J.~Dixon,
  JHEP {\bf 0306}, 028 (2003)
  [arXiv:hep-ph/0304168].

\bibitem{TwoLoopSplitting}
  Z.~Bern, L.~J.~Dixon and D.~A.~Kosower,
  JHEP {\bf 0408}, 012 (2004)
  [arXiv:hep-ph/0404293].

\bibitem{DeFreitas:2004tk}
  A.~De Freitas and Z.~Bern,
  JHEP {\bf 0409}, 039 (2004)
  [arXiv:hep-ph/0409007].

\bibitem{TwoLoopBasis}
J.~Gluza, K.~Kajda and D.~A.~Kosower,
Phys.\ Rev.\  D {\bf 83}, 045012 (2011)
[arXiv:1009.0472 [hep-th]].

\bibitem{Octacut}
E.~I.~Buchbinder and F.~Cachazo,
JHEP {\bf 0511}, 036 (2005)
[arXiv:hep-th/0506126].

\bibitem{IBP}
F.~V.~Tkachov,
Phys.\ Lett.\ B {\bf 100}, 65 (1981); \\
K.~G.~Chetyrkin and F.~V.~Tkachov,
Nucl.\ Phys.\ B {\bf 192}, 159 (1981).

\bibitem{Laporta}
S.~Laporta,
  Phys.\ Lett.\  B {\bf 504}, 188 (2001)
  [hep-ph/0102032].
S.~Laporta,
Int.\ J.\ Mod.\ Phys.\ A {\bf 15}, 5087 (2000)
[hep-ph/0102033].

\bibitem{GehrmannRemiddi}
T.~Gehrmann and E.~Remiddi,
Nucl.\ Phys.\  B {\bf 580}, 485 (2000)
[hep-ph/9912329].

\bibitem{LIdependent}
R.~N.~Lee,
  JHEP {\bf 0807}, 031 (2008)
  [arXiv:0804.3008 [hep-ph]].

\bibitem{AIR}
C.~Anastasiou and A.~Lazopoulos,
JHEP {\bf 0407}, 046 (2004)
[hep-ph/0404258].

\bibitem{FIRE}
A.~V.~Smirnov,
JHEP {\bf 0810}, 107 (2008)
[0807.3243 [hep-ph]].

\bibitem{Reduze}
C.~Studerus,
Comput.\ Phys.\ Commun.\  {\bf 181}, 1293 (2010)
[arXiv:0912.2546 [physics.comp-ph]].

\bibitem{SmirnovPetukhov}
A.~V.~Smirnov and A.~V.~Petukhov,
arXiv:1004.4199 [hep-th]. 

\bibitem{IntegralReductions}
L.~M.~Brown and R.~P.~Feynman,
Phys.\ Rev.\  {\bf 85}, 231 (1952);\\
L.~M.~Brown, Nuovo Cim.\ {\bf 21}, 3878 (1961);\\
%
B.~Petersson,  J. Math. Phys.\ {\bf 6}, 1955 (1965);\\
%
G.~K\"all\'en and J.~S.~Toll, J. Math.\ Phys.\ {\bf 6}, 299 (1965);\\
%
D.~B.~Melrose,
Nuovo Cim.\  {\bf 40}, 181 (1965);\\
%
G.~Passarino and M.~J.~G.~Veltman,
Nucl.\ Phys.\  B {\bf 160}, 151 (1979);\\
%
W.~L.~van Neerven and J.~A.~M.~Vermaseren,
Phys.\ Lett.\  B {\bf 137}, 241 (1984);\\
G.~J.~van Oldenborgh and J.~A.~M.~Vermaseren,
Z.\ Phys.\  C {\bf 46}, 425 (1990).

\bibitem{RSV}
  R.~Roiban, M.~Spradlin and A.~Volovich,
  Phys.\ Rev.\  D {\bf 70}, 026009 (2004)
  [arXiv:hep-th/0403190].

\bibitem{WittenTwistorString}
E.~Witten,
Commun.\ Math.\ Phys.\  {\bf 252}, 189 (2004)
[arXiv:hep-th/0312171].

\bibitem{Vergu}
C.~Vergu,
Phys.\ Rev.\  D {\bf 75}, 025028 (2007)
[arXiv:hep-th/0612250].

\bibitem{Bullimore:2009cb}
M.~Bullimore, L.~J.~Mason and D.~Skinner,
JHEP {\bf 1003}, 070 (2010)
[arXiv:0912.0539 [hep-th]].

\bibitem{Shabat}
B.~V.~Shabat,
Introduction to Complex Analysis, Part II: Functions of Several Variables.
American Mathematical Society, Vol. 110, 1992.

\bibitem{Elvang:2008na}
H.~Elvang, D.~Z.~Freedman and M.~Kiermaier,
JHEP {\bf 0904}, 009 (2009)
[arXiv:0808.1720 [hep-th]].

\bibitem{Bern:2009xq}
Z.~Bern, J.~J.~M.~Carrasco, H.~Ita, H.~Johansson and R.~Roiban,
Phys.\ Rev.\  D {\bf 80}, 065029 (2009)
[arXiv:0903.5348 [hep-th]].

\bibitem{DixonPrivate}
L.~J.~Dixon, private communication.

\bibitem{MastroliaOssola}
P.~Mastrolia and G.~Ossola,
arXiv:1107.6041 [hep-ph].

\bibitem{LarsenSixPoint} 
K.~J.~Larsen,
arXiv:1205.0297 [hep-th].

\end{thebibliography}
\end{document}